\documentclass[final,3p,twocolumn]{elsarticle}

\usepackage{amssymb}
\usepackage{amsthm,amsmath}
\usepackage{caption}
\usepackage{lmodern}
\usepackage{lineno}
\journal{arXiv}

\begin{document}
\begin{frontmatter}


\title{Effects of Cr content on ion-irradiation hardening of FeCrAl ODS ferritic steels with 9 wt\% Al}


\author[1,2]{Zhexian Zhang\corref{cor1}}
\ead{zzhan124@utk.edu}
\author[1,2]{Siwei Chen}
\author[1]{Kiyohiro Yabuuchi}

\affiliation[1]{organization={Institute of Advanced Energy, Kyoto University},
            addressline={Gokasho}, 
            city={Uji},
            postcode={611-0011}, 
            state={Kyoto},
            country={Japan}}
            
 \affiliation[2]{organization={Department of Nuclear Engineering, University of Tennessee},
             city={Knoxville},
             postcode={37996},
             state={TN},
             country={USA}}

\cortext[cor1]{corresponding author}

\begin{abstract}
FeCrAl ODS steels for accident tolerant fuel claddings are designed to bear high-Cr and Al for enhancing oxidation resistance. In this study, we investigated the effects of Cr content on ion-irradiation hardening of three ODS ferritic steels with different Cr contents added with 9 wt\% Al, Fe12Cr9Al (SP12), Fe15Cr9Al (SP13), and Fe18Cr9Al (SP14). The specimens were irradiated with 6.4MeV Fe\textsuperscript{3+} at 300 \textdegree C to nominal 3 dpa. The irradiation hardening was measured by nanoindentation method, and the Nix-Gao plots were used to evaluate the bulk-equivalent hardness. The results showed that the irradiation hardening decreased with increasing Cr content. The reason is due to the growth of dislocation loops hindered by solute Cr atoms. TEM observations showed both $\langle 100\rangle$ and $1/2\langle 111\rangle$ dislocation loops existed in the irradiated area. The irradiation hardening was estimated by dispersed barrier hardening (DBH) model with dislocation loops.
\end{abstract}

%

\begin{keyword}


FeCrAl ODS steel \sep ion-irradiation\sep nanoindentation\sep dislocation loops\sep Cr content 
\end{keyword}

\end{frontmatter}


\section{Introduction} \label{sec:1}
Accident tolerant fuel (ATF) cladding has been proposed to enhance the safety of light water reactors (LWR) under design-basis and beyond-design-basis accident scenarios, aiming to delay the onset of detrimental oxidation and chemical interaction processes with hot water\citep{zinkleAccidentTolerantFuels2014,terraniAccidentTolerantFuel2018}. FeCrAl ODS ferritic steels have been considered as a promising candidate because of its excellent mechanical properties and good oxidation resistance at high temperatures\citep{kimuraDevelopmentFuelClad2003,kimuraSuperODSSteels2009,kimuraDevelopmentAddedHighCr2011}. During operation, the FeCrAl claddings experience neutron irradiation at operating temperature around 280 to 300\textdegree C, which induces generation of defects such as dislocation loops, voids, precipitates and so on. These defects may lead to irradiation embrittlement, for example, shift the ductile brittle transition temperature (DBTT) greatly over room temperature\citep{maloyViabilityThinWall2016}. This degradation could be reflected by the irradiation hardening accompanied by lowering ductility\citep{maloyCharacterizationComparativeAnalysis2016}. 

The FeCrAl ODS steels are designed with different Cr ($<~$20wt\%) and Al ($<~$10wt\%) contents. The upper-limit content of Cr can be due to the $\alpha$-$\alpha^\prime$ phase separation during service at elevated temperatures through spinodal and nucleation-growth mechanism according to the miscibility gap\citep{kobayashiMapping475Embrittlement2010,chenAgehardeningSusceptibilityHighCr2015}. On the other hand, the Al content is limited to below 10wt\%, which is solubility limit of Al in Fe at around 300 \textdegree C\citep{majiMicrostructuralStabilityIntermetallic2021}. High Al will also induce the formation of Al-Ti enriched $\beta^\prime$ precipitates, which causes embrittlement during early aging at 475\textdegree C\citep{sangEarlystageThermalAgeing2020,kimuraTwofoldAgehardeningMechanism2023}.

So far, irradiation effects on FeCrAl ODS steels are mainly studied for the steels with Al around 3$~$6 wt\%\citep{kondoIonIrradiationEffects2018,songRadiationResponseODS2018,songInsightsHardeningPlastically2022,gaoDoseDependenceIon2020,gettoFrictionStirWelding2022,gettoUnderstandingRadiationEffects2022}. However, high Al could shift the miscibility boundary of $\alpha-\alpha^\prime$ phase separation to a higher Cr content[8,19]\citep{kobayashiMapping475Embrittlement2010,ejenstamMicrostructuralStabilityFe2015}, consequently hinders the $\alpha^\prime$ precipitation hardening. Moreover, corrosion tests showed that the addition of Al (2$~$4wt\%) efficiently reduced the weight gains in supercritical pressurized water (SCPW)\citep{kimuraDevelopmentAddedHighCr2011}. With these merits of Al, it is benefit to investigate ion-irradiation hardening of FeCrAl ODS steels with rather higher Al content. Zhou et al\citep{zhouEffectAluminumContent2019} studied the Al effect on formation of the dislocation loops in non-ODS FeCrAl steels. The results showed Al can effectively suppress the growth of the dislocation loops due to enhanced pinning effect of Al atoms.

Another issue of FeCrAl alloys is the effect of Cr content on irradiation hardening. As is well known, Cr improves the corrosion resistance by forming protective $Cr_2O_3$ layers. However, the irradiation embrittlement could be accelerated by increasing Cr content which causes $\alpha-\alpha^\prime$ separation by radiation enhanced diffusion\citep{reeseDoseRateDependence2018}. The $\alpha^\prime$ precipitation in FeCrAl steels has been studied and proved susceptible to alloy composite\citep{edmondsonIrradiationenhancedPrecipitationModel2016}, temperature\citep{masseyPostIrradiationExamination2019} and grain morphology\citep{zhangInfluenceWeldingNeutron2019,maoImprovedIrradiationResistance2022}, and the precipitation kinetics under irradiation follows typical Lifshitz-Slyozov-Wagner (LSW) theory\citep{fieldPrecipitationNeutronIrradiated2018}. Both the ion-irradiation and neutron irradiation showed Cr-dependence of irradiation hardening in FeCrAl steels\citep{fieldRadiationToleranceNeutronirradiated2015,aydoganMicrostructureMechanicalProperties2018}.

To fill the gap of research in high-Al FeCrAl ODS steels, we investigated the FeCrAl ODS ferritic steels with 9wt\% Al which is close to its up-limit. Particularly to understand the Cr-dependent irradiation hardening, steels with different concentration of Cr (12, 15, 18wt\%) were studied. These steels were subjected to self-ion irradiation to nominal 3 dpa at 300\textdegree C. The hardness of steels was measured by nanoindentation, and microstructures were observed by TEM. The Cr dependent hardening in FeCrAl ODS steels with high Al concentrations was discussed. 

\section{Experimental}\label{sec:2}
Three FeCrAl ODS ferritic steels, of which the chemical compositions are shown in Table \ref{tab:composition}. These steels were produced by mechanical alloying in Ar atmosphere. The mixed powder was encapsulated into a steel capsule, hot-extruded at 1150\textdegree C and followed by the final step of the heat treatment at 1150\textdegree C for 1hr followed by air cooling. The detail of production procedure is available in a recent review\citep{ukaiAlloyDesignCharacterization2023}. Specimens for irradiation were cut so that the irradiation surface is parallel to the extrusion direction. The specimens were grinded with SiC paper and polished with 3$\mu m$, 1$\mu m$, 0.25$\mu m$ diamond powders subsequently. Final surface treatment included electrical polishing in 5\% $HClO_4$ and 95\% $CH_3COOH$ for 60 sec, and low-energy Ar ion milling for 10 min. 

\begin{table*}[bt] \label{tab:1}
\centering 
 \caption{Chemical compositions of FeCrAl ODS ferritic steels (wt\%, Bal. Fe)} 
\begin{tabular}{l c c c c c c c c c}
\hline 
 & ID & Cr & Al & Ti & Y & C & O & N & Ar \\ 
\hline 
Fe12Cr9Al & SP12 & 11.93 & 8.65 & 0.53 & 0.38 & 0.029 & 0.22 & 0.003 & 0.006 \\ 
Fe15Cr9Al & SP13 & 14.25 & 8.4 & 0.51 & 0.38 & 0.03 & 0.22 & 0.003 & 0.006 \\  
Fe18Cr9Al & SP14 & 16.63 & 8.09 & 0.49 & 0.37 & 0.032 & 0.22 & 0.003 & 0.006 \\ 
\hline 
\end{tabular} 
\end{table*}

\begin{table*}[bt]
\centering
 \caption{The dislocation loops and DBH hardening after irradiation.}\label{tab:loops}
\begin{tabular}{l c c c c}
\hline 
 & \multicolumn{2}{c}{Dislocation loops} & \multicolumn{2}{c}{Hardening (GPa)} \\ 
 & Diameter (nm) & Density ($\times 10^{22} /m^3$) & Measured & Calculated \\ 
 \hline 
Fe12Cr9Al (SP12) & $11.6\pm2.0$ & $2.26\pm0.34$ & $1.45\pm0.27$ & $1.28$ \\ 
Fe15Cr9Al (SP13) & $12.0\pm1.0$ & $1.84\pm0.85$ & $1.35\pm0.33$ & $1.18$ \\ 
\hline 
\end{tabular} 
\end{table*}

\begin{table*}[bt]
\centering
 \caption{The bulk-equivalent nano-hardness (GPa) evaluated by different range of Nix-Gao plots.}\label{tab:ng}
\begin{tabular}{c c c c}
\hline 
Range of Nix-Gao plots & Fe12Cr9Al (SP12) irradiated & Fe15Cr9Al (SP13) irradiated & Fe18Cr9Al (SP14) irradiated \\ 
\hline 
100nm-250nm & $6.23\pm0.27$ & $6.22\pm0.35$ & $5.72\pm0.31$ \\  
100nm-300nm & $6.21\pm0.23$ & $6.19\pm0.31$ & $5.80\pm0.26$ \\ 
100nm-350nm & $6.16\pm0.21$ & $6.16\pm0.27$ & $5.83\pm0.23$ \\ 
150nm-250nm & $6.22\pm0.29$ & $6.14\pm0.38$ & $5.95\pm0.32$ \\ 
200nm-300nm & $6.12\pm0.28$ & $6.08\pm0.27$ & $6.02\pm0.38$ \\ 
200nm-350nm & $5.97\pm0.29$ & $6.04\pm0.30$ & $5.95\pm0.43$ \\ 
\hline 
\end{tabular} 
\end{table*}

\begin{figure}[btp] 
\centering
\includegraphics[width=\linewidth]{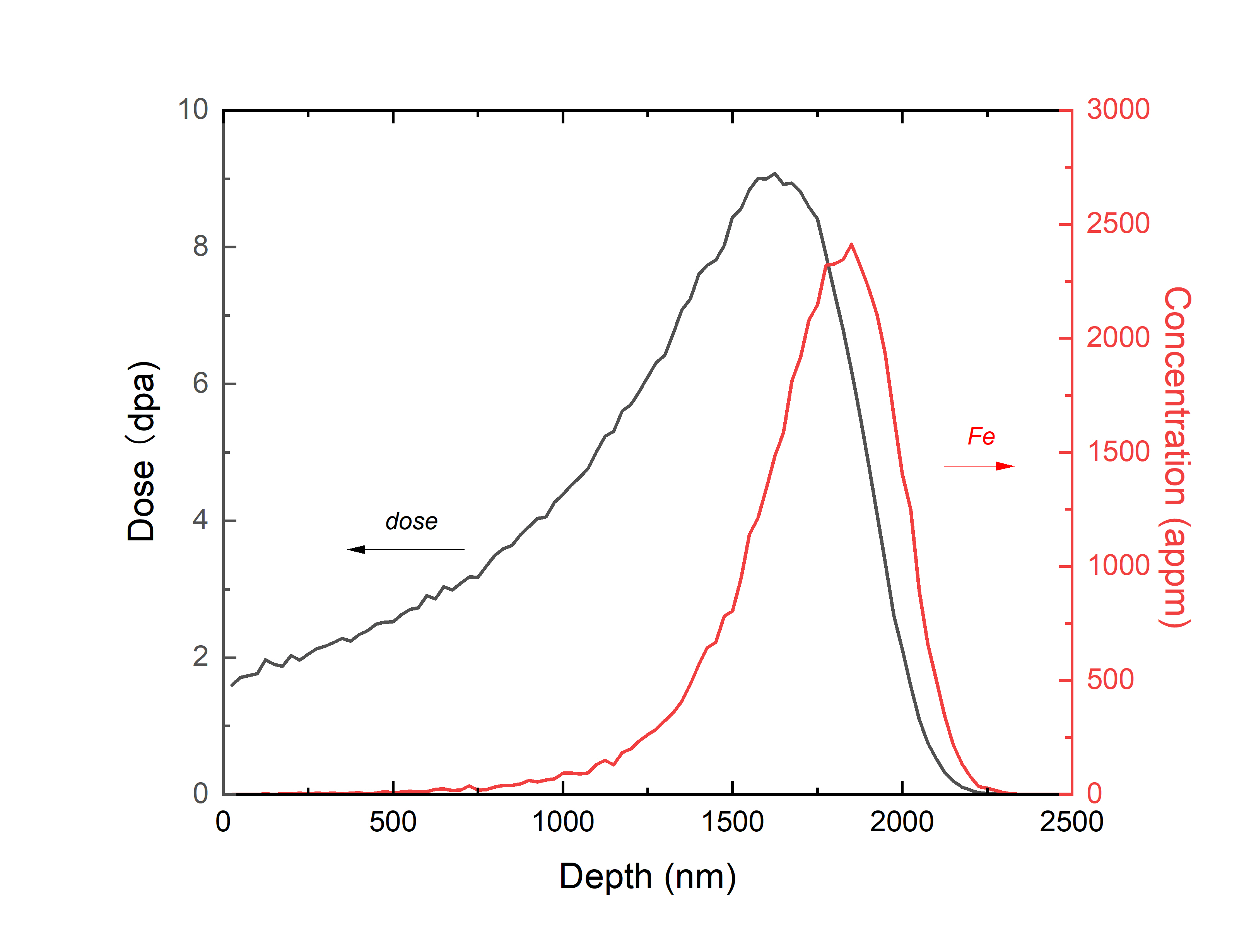}
\caption{Irradiation dose and implanted Fe ions distribution calculated by SRIM. 3 dpa at 600 nm was selected as a nominal value.}\label{fig:srim}
\end{figure}

\begin{figure*}[btp] 
\centering
\includegraphics[width=0.6\textwidth]{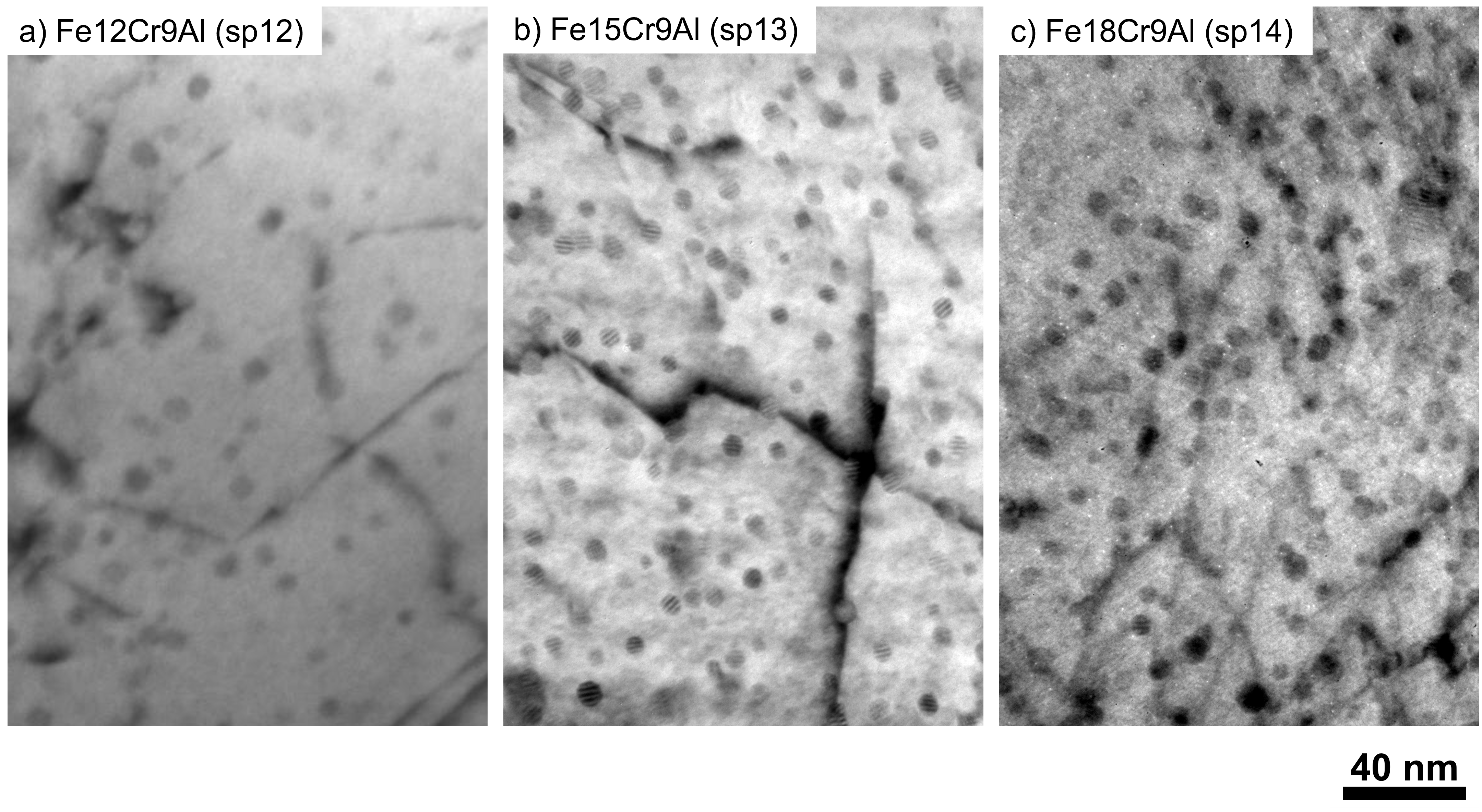}
\caption{The oxides morphology in a) Fe12Cr9Al, b) Fe15Cr9Al, c) Fe18Cr9Al ODS steels before irradiation.}\label{fig:oxides}
\end{figure*}

\begin{figure}[btp] 
\centering
\includegraphics[width=\linewidth]{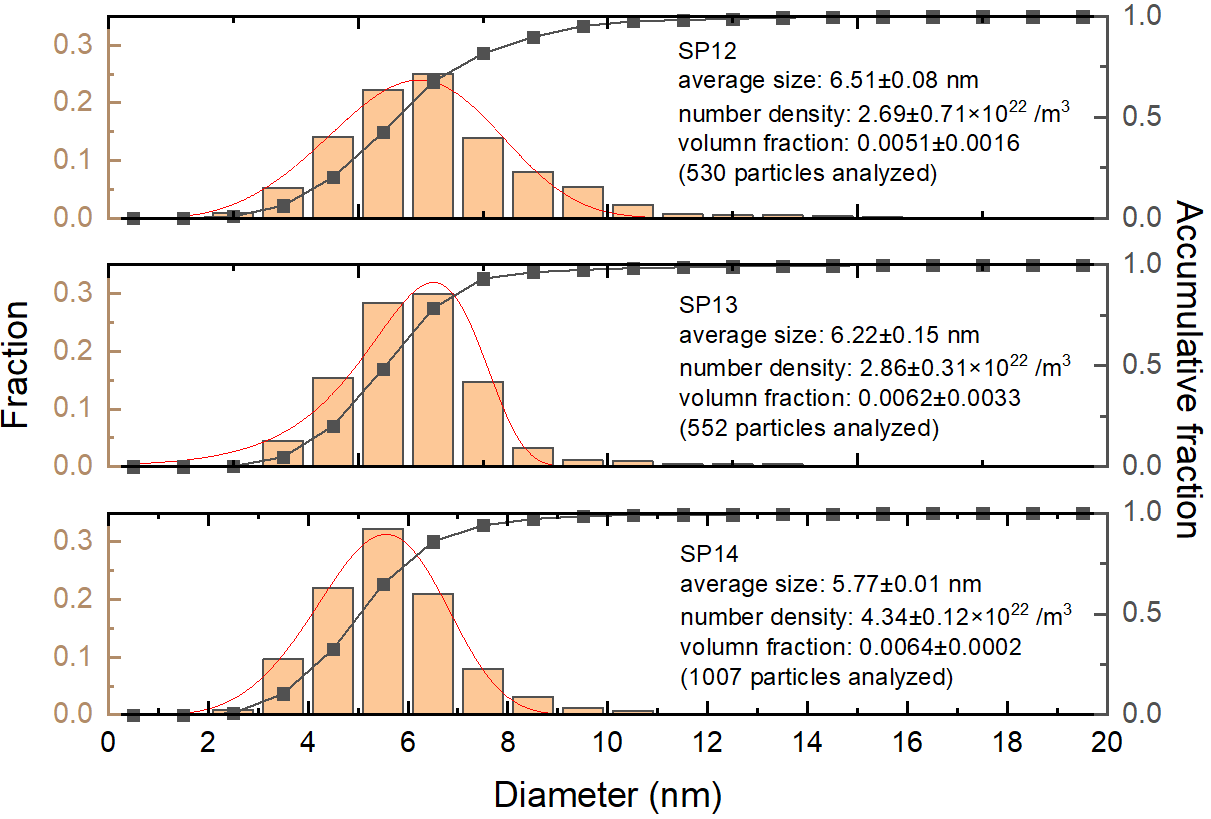}
\caption{The size distribution of oxides in unirradiated Fe12Cr9Al (SP12), Fe15Cr9Al (SP13), and Fe18Cr9Al (SP14) ODS steels. The red lines are fitted results by Weibull distribution. The volume fractions were obtained by an average of observed three areas.}\label{fig:fraction}
\end{figure}

\begin{figure}[btp] 
\centering
\includegraphics[width=\linewidth]{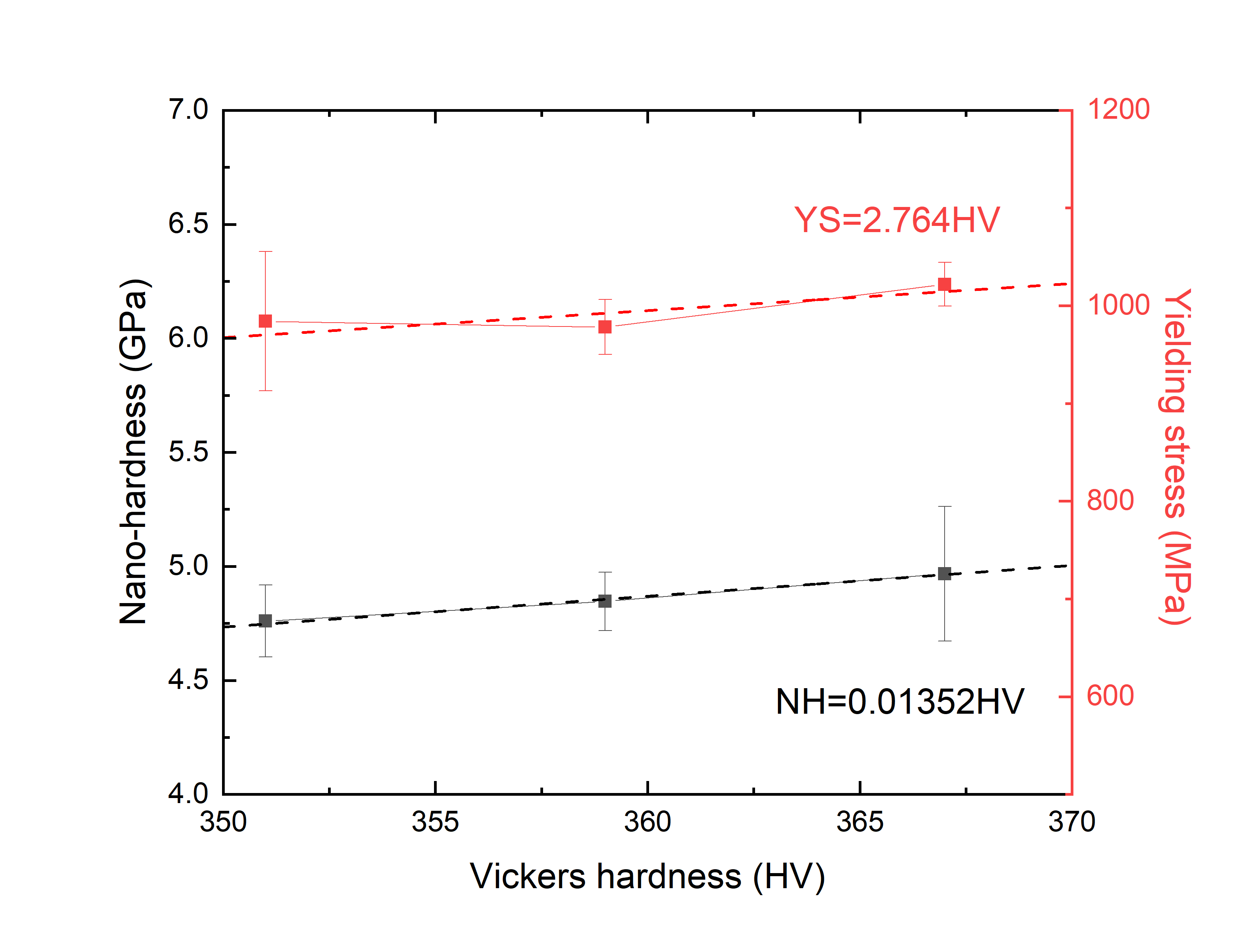}
\caption{The relationship between nano-hardness (NH), Vickers hardness (HV) and $\sigma_{0.2}$ yielding stress (YS) in the three ODS ferritic steels before irradiation. The nano hardness was obtained by Nix-Gao methods.  The intercept was preset as zero for linear fitting.}\label{fig:fitting}
\end{figure}

\begin{figure*}[btp] 
\centering
\includegraphics[width=\textwidth]{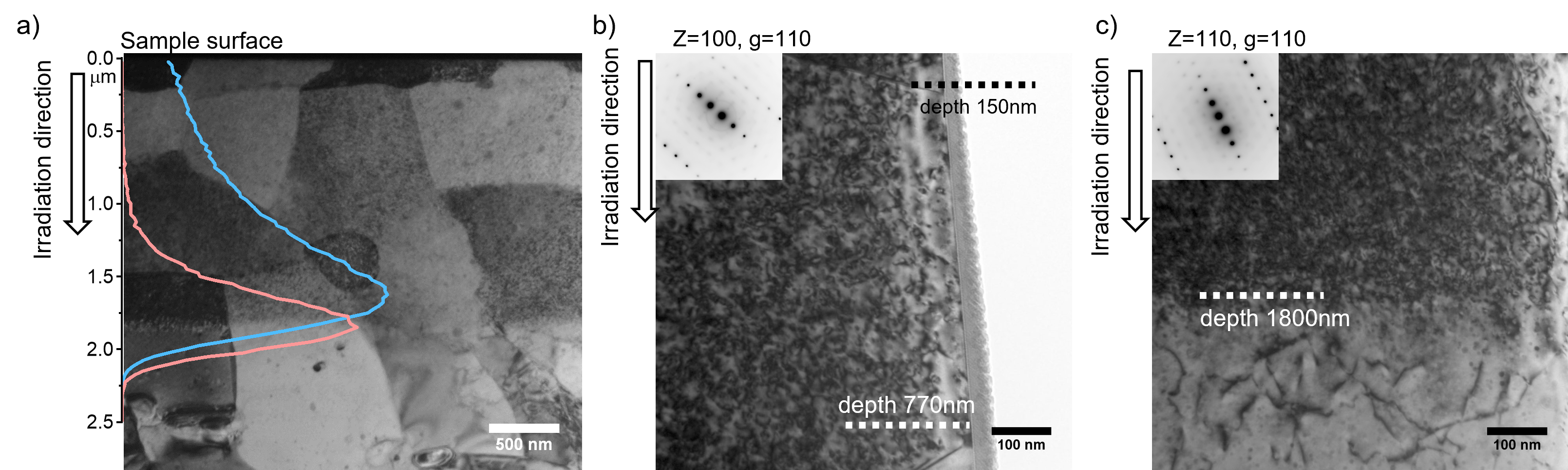}
\caption{The relationship between nano-hardness (NH), Vickers hardness (HV) and $\sigma_{0.2}$ yielding stress (YS) in the three ODS ferritic steels before irradiation. The nano hardness was obtained by Nix-Gao methods.  The intercept was preset as zero for linear fitting.}\label{fig:sp12}
\end{figure*}

\begin{figure*}[btp] 
\centering
\includegraphics[width=0.6\textwidth]{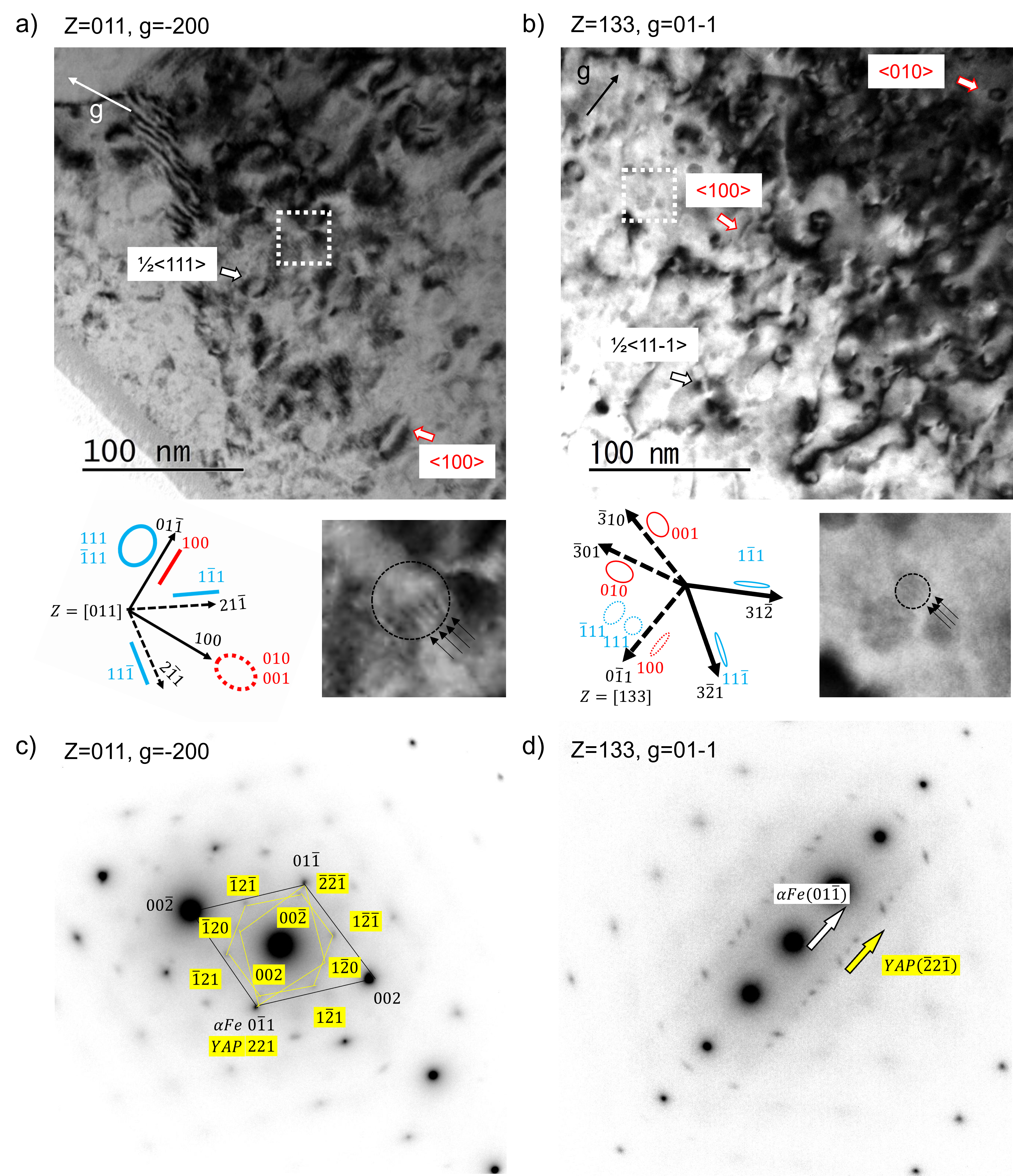}
\caption{Microstructures of irradiated Fe15Cr9Al ODS steel (SP13), a) g=200, b) g=110. Dislocation loops projections were shown in each zone axis. The $Moir\acute{e}$ fringes of oxides indicate crystal structure was kept under irradiation. The diffraction pattern of the irradiated area along c) [011] direction and d) [133] direction.}\label{fig:sp13}
\end{figure*}

\begin{figure*}[btp] 
\centering
\includegraphics[width=0.9\textwidth]{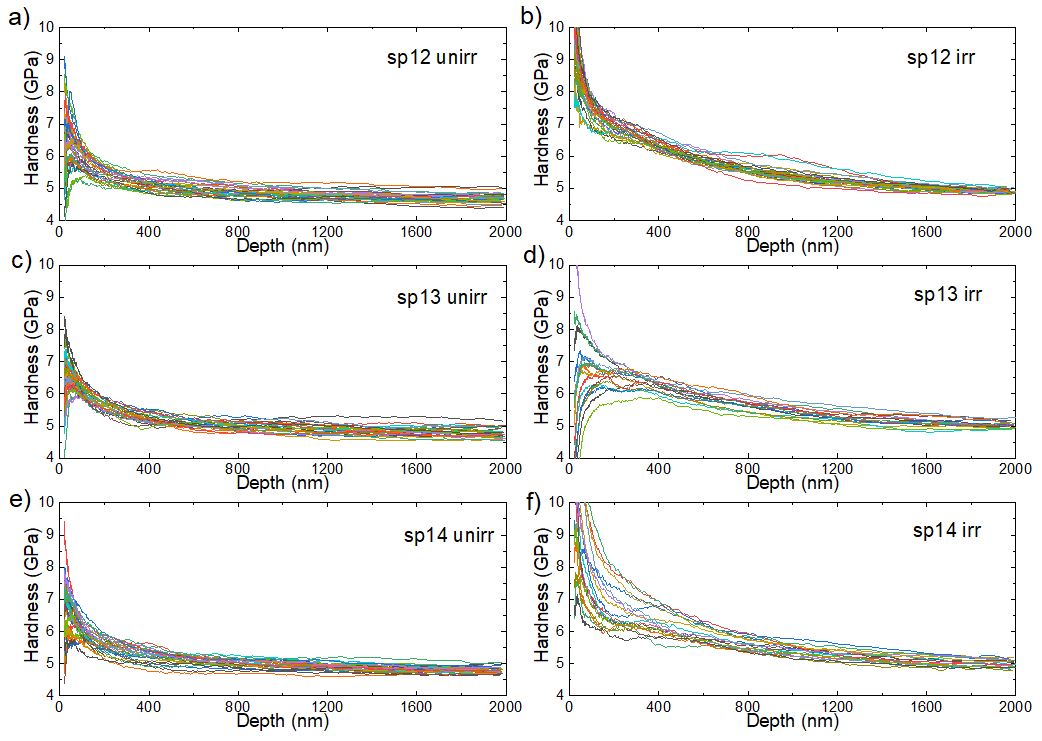}
\caption{The nano-indentation hardness profile of a) unirradiated and b) irradiated Fe12Cr9Al (SP12), c) unirradiated and d) irradiated Fe15Cr9Al (SP13), e) unirradiated and f) irradiated Fe18Cr9Al (SP14).}\label{fig:hd}
\end{figure*}

Ion irradiation was performed with Dual-beam Facility for Energy Science and Technology (DuET) in Kyoto University. DuET is composed of a 1.7 MV tandem accelerator for heavy ion irradiation and a 1.0MV single accelerator for helium implantation. Details of the construction can be found in Ref\citep{kohyamaNewMultipleBeams2000}. The ion source of $Fe_2O_3$ was ionized by an 860 ionizer in Cesium reservoir. The ionized charged particle, $FeO^-$, was focused by Einzel lens and deflected by an injector magnet to accelerator. In the tandem accelerator, the $FeO^-$ beam was first accelerated to 1.7 MeV halfway, then stripped into 1.32 MeV $Fe^{3+}$ and 0.38 MeV oxygen ions. The $Fe^{3+}$ was accelerated again in the other half of the accelerator with the total energy of $Fe^{3+}$ becoming 6.4MeV. A magnet accelerator was used thereafter to screen out ions other than 6.4 MeV $Fe^{3+}$. Beam position was finally adjusted by a steerer, and beam current was measured by array of 21 Faraday cups. The vacuum of the target chamber was $5\times 10^{-5}$ Pa. The beam irradiated on specimen was raster-scanned horizontally 1000 Hz and vertically 300 Hz. The temperature of irradiated sample surface was monitored by infrared thermography during irradiation. The emissivity of materials was calculated by pre-calibration with a K-type thermocouple attached to the specimen surface at various temperatures before irradiation.

The irradiation damage and implanted ion distribution (Fig.\ref{fig:srim}) were estimated by SRIM\citep{zieglerSRIMStoppingRange2010} using the “Ion distribution and quick calculation of damage” method. The displacement threshold energy(TDE) was selected by 40 eV\citep{astme521-96e1StandardPracticeNeutron2009} for iron-based steels. The value at the depth of 600 nm was chosen as the nominal irradiation dose because there is barely interference of the implanted irons at this range. The samples were irradiated to nominal 3 dpa at $300\pm 10 \textdegree C$ with the beam flux of $1.1\pm 10^{12} ions/cm^2/s$, which is corresponding to $3.3\pm 10^{-4} dpa/s$. The peak dose is located around 1.6 $\mu m$. The maximum damage distribution is about 2.3 $\mu m$. 

Nano-hardness was measured by G200 nanoindentation with a Berkovich tip using continuous stiffness measurement (CSM) method. The oscillation amplitude was 2 nm, and the frequency was 45 Hz. The strain rate is constant 0.05 $s^{-1}$, which was achieved by loading rate control\citep{lucasIndentationPowerlawCreep1999}. The area function was calibrated on fused silica (E=72.0GPa, $\nu =0.17$). Over 20 tests were performed on each sample. The tests were indented to maximum 2 $\mu m$ depth and held for 10 sec at the maximum loading. 

Tensile tests were performed on the tensile machine (INTESCO Co., Ltd) with a load cell of 0.5 kN. The geometries of a dog-born sheet type of miniature tensile specimens measured gauge-length=5 mm, width=1.2 mm and thickness=0.25 mm with the gripping area on both sides of $4\times 4mm^2$. The displacement speed was 0.2 mm/min resulting in an initial strain rate of $6.67\times 10^{-4} /s$. The yield stress was defined as 0.2\% off-set flow stress. Three tests for each material were done at room temperature. Micro-Vickers hardness was measured at room temperature by a hardness tester, HMV-2T (Shimadzu Corp.), with 2 kg load and holding time for 10 sec. 

Microstructures of specimens were observed by Jeol 2010 TEM with a side-mount CCD camera. The cross-section of irradiated specimens was fabricated by focused ion beam machining (FIB, Hitachi FB2200). TEM specimens were flashing-polished in 5\% HClO4 and 95\% CH3OH at -30 \textdegree C and 30 V as final thinning. The images were taken under two-beam bright field condition, with (110) $\alpha$Fe diffraction plane excited. The thickness of TEM specimens was measured by extinction fringes along specimen margins or inclined non-twinned large angle grain boundaries under g3g weak beam central dark field condition. 

\section{Results}\label{sec:3}
\subsection{The oxide morphology and mechanical properties before irradiation}\label{subsec:3.1}
The oxides in Al-added FeCrAl ODS steels are mainly Y-Al-O types. The lattice image analysis by Oono et al showed that most of the oxides were $Y_4Al_2O_9$ (YAP) and $YAlO_3$ (YAM)\citep{oonoGrowthOxideParticles2017}. The morphology of dispersoids before irradiation were shown in Fig.\ref{fig:oxides}. Both dislocation lines and oxides exist in the materials. The dislocations are straight lines suggesting that the tensions to the dislocations are fully relieved during the heat treatment after mechanical alloying. Some interactions between dislocation lines and oxides can be observed but there is still no curved dislocation line around oxide particles.

Fig.\ref{fig:fraction} shows the size distribution of oxide particle diameters in the three as-received FeCrAl ODS steels. The diameters are mainly located in a rather small range between 4~8 nm, which means the size of oxides are generally homogeneous. The average diameter of the oxides tends to increase with decreasing Cr content. The Fe18Cr9Al has the highest number densities of oxides in the three steels. The estimated volume fractions are 0.51\%, 0.62\% and 0.64\% for Fe12Cr9Al, Fe15Cr9Al, and Fe18Cr9Al, respectively.

The bulk equivalent hardness from nanoindentation tests were estimated by Nix-Gao method\citep{nixIndentationSizeEffects1998} in equation \ref{eq:1}, where $H$ is the measured hardness, $H0$ is the bulk equivalent hardness, $h*$ is a length scale indicator, $h$ is the indentation depth. 

\begin{equation}\label{eq:1}
H^{2}=H_{0}^{2}\left( 1+\frac{h^{*}}{h} \right)
\end{equation}

Bulk-equivalent nano-hardness of unirradiated materials were estimated by data between 100 to 2000nm. Relationships among nano-hardness (NH), Vickers hardness (HV) and yielding stress (YS, $\sigma_{0.2}$) were shown in Fig.\ref{fig:fitting}. The NH=0.01352HV is similar to the previous obtained results\citep{yabuuchiEvaluationIrradiationHardening2014}. However, the YS=2.764HV is somewhat different from the common acknowledged 3 times rules\citep{busbyRelationshipHardnessYield2005}.

\subsection{The microstructure evolution after irradiation}\label{subsec:3.2}
Fig.\ref{fig:sp12} shows the cross-section of irradiation damaged Fe12Cr9Al (SP12). The overall damaged region was compared to the SRIM calculated dose profile (Fig.\ref{fig:sp12}a). The dislocation loops distributed to the maximum depth of 1.8 $\mu m$, which exceeded the damage peak, but was shallower than the maximum irradiation range calculated by SRIM. Fig.\ref{fig:sp12}b shows the microstructures in the region between 150~770 nm, which was selected as the area for TEM analysis. In this region, dislocation loops coexisted with oxide particles and dislocation lines. Fig.\ref{fig:sp12}c shows the transition area of the irradiated region. In this region, dislocation loops disappeared and only dislocation lines and oxides existed in the unirradiated area.

Fig.\ref{fig:sp13} shows two different areas around 600 nm depth of the irradiated Fe15Cr9Al (SP13) under zone axis [011] and [133], with g vectors equals to -200 and 01-1 respectively. The projection images of $1/2<111>$ and $<100>$ dislocation loops with different habit planes under different zone axis were schematically shown. The solid circles are the loops appearing with contrast and the dashed circles are the loops being extinct at this diffraction condition. Both Fig.\ref{fig:sp13}a-b proved the existence of the $1/2<111>$ and $<100>$ type dislocation loops and the oxide particles in the irradiated region. The $Moir\acute{e}$ fringes indicated that the oxides were not amorphized. The diffraction patterns in Fig.\ref{fig:sp13}c-d showed that the oxides were mainly YAP, with $\alpha$Fe(0-11) corresponding to YAP(221) because they have similar plane spacing. Fig.\ref{fig:sp13}c also showed another relationship of $\alpha$Fe(001)//YAP(1-20) and $\alpha$Fe[100]//YAP[210].

Song et al\citep{songAssessmentPhaseStability2020a} studied the stability of oxide particles in ODS steels under ion-irradiation at 200 \textdegree C based on the observation of $Moir\acute{e}$ fringes, and reported that the Y-Al-O oxides tended to become amorphous under irradiation up to ~8 dpa and dissolved thereafter. They suggested that the radiation tolerance of the oxide particles depended on both the irradiation temperature and damage rate as well as total damage (dpa). In this study, under the ion-irradiation at 300 \textdegree C and almost the same damage rate, the oxides were not amorphized, which means the Y-Al-O system oxides are relatively stable at this irradiation condition. It might be also owing to the smaller dose used in this experiment compared to the study of Song et al\citep{songAssessmentPhaseStability2020a}.

The average diameter and number density of dislocation loops were summarized in Table \ref{tab:loops}, where the density of dislocation loops was adjusted according to the invisible criterion in TEM two beam diffraction condition. The diameters of dislocation loops in SP12 and SP13 are $11.6\pm 2.0nm$ and $12.0\pm 1.0nm$, and the number densities are $2.26\pm 0.34\times 10^{22}/m^3$ and $1.84\pm 0.85\times 10^{22}/m^3$, respectively.

\subsection{The nano-indentation hardness.}\label{subsec:3.3}
The profiles of nanoindentation hardness of irradiated and unirradiated FeCrAl ODS steels are shown in Fig.\ref{fig:hd}. The features of the profiles include: 1) the hardness decreases gradually along with indentation depth because of the indentation size effect, 2) a reverse size effect in some tests before ~100nm occurred possibly because of initial contact variance from surface roughness, grain boundaries and precipitates, but as the indent went deeper, the normal size effect became dominant. 3) the irradiated samples have elevated hardness than the unirradiated ones indicating irradiation hardening. 4) the hardening disappeared at an indentation depth around 2000 nm where the plastic deformation zone in the irradiated region took small fraction compared to the unirradiated region. 

\begin{figure}[btp] 
\centering
\includegraphics[width=0.8\linewidth]{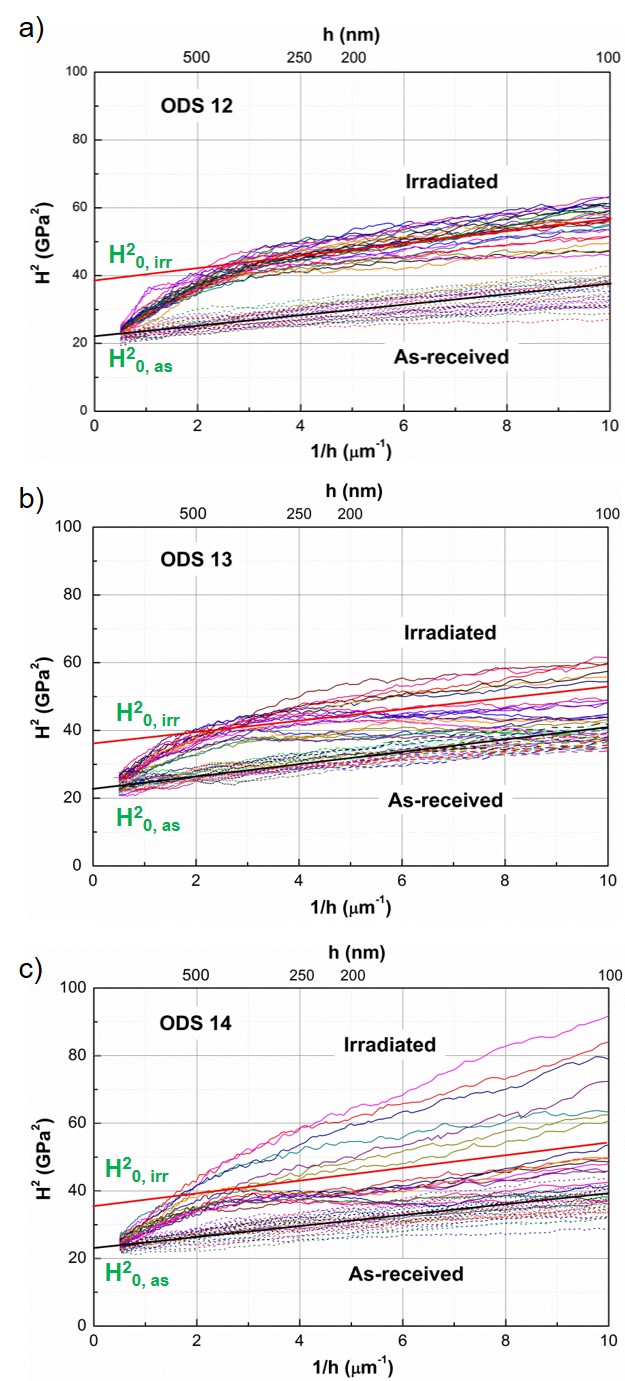}
\caption{Nix-Gao plots of the irradiated and unirradiated a) Fe12Cr9Al ODS steel (SP12), b) Fe15Cr9Al ODS steel (SP12) and c) Fe18Cr9Al ODS steel (SP14). Note that the manually drawn lines are only an indicator to the shoulder for linear fitting and do not correspond to the actual bulk equivalent hardness.}\label{fig:NG}
\end{figure}

The Nix-Gao hardness plots of both irradiated and unirradiated steels are shown in Fig.\ref{fig:NG}. Before irradiation, the Nix-Gao plots were linear through the entire exhibited range. In contrast, there is a clear shoulder of irradiated plots in the depth region lower than 300 nm, which reflects the presence of layered structure in the intended region. The bulk-equivalent hardness was estimated by the linear fitting of the data between 100 nm~2000 nm in the unirradiated specimens. The hardness after irradiation was evaluated by data with different ranges between 100 nm~350 nm of Nix-Gao plots. The depth of damage limit at 1800 nm corresponds to about 5 times of the turning point at 350 nm. 

The depth range selection for hardness evaluation of irradiation is not only dependent on the linear correlation. This is because the profile of Nix-Gao plots is usually not absolutely linear\citep{hausildMethodologicalCommentNanoindentation2021}. Moreover, the surface roughness\citep{walter3D2DFinite2009} and pile up\citep{zhuAccurateEvaluationBulk2022} will affect the measured hardness not only increasing scattering but change the linear relationship. The bulk-equivalent nano-hardness evaluated by different range of Nix-Gao plots were summarized in Table \ref{tab:ng}. For Fe12Cr9Al and Fe15Cr9Al, as the selected range goes deeper, the evaluated hardness keeps decreasing. Oppositely, the evaluated hardness of Fe18Cr9Al increases with selected depth until around 6 GPa. These are typical phenomenon in irradiated materials, where the surfaces are inhomogeneous either because of irradiation induced defects or surface roughness.

\begin{figure}[btp] 
\centering
\includegraphics[width=\linewidth]{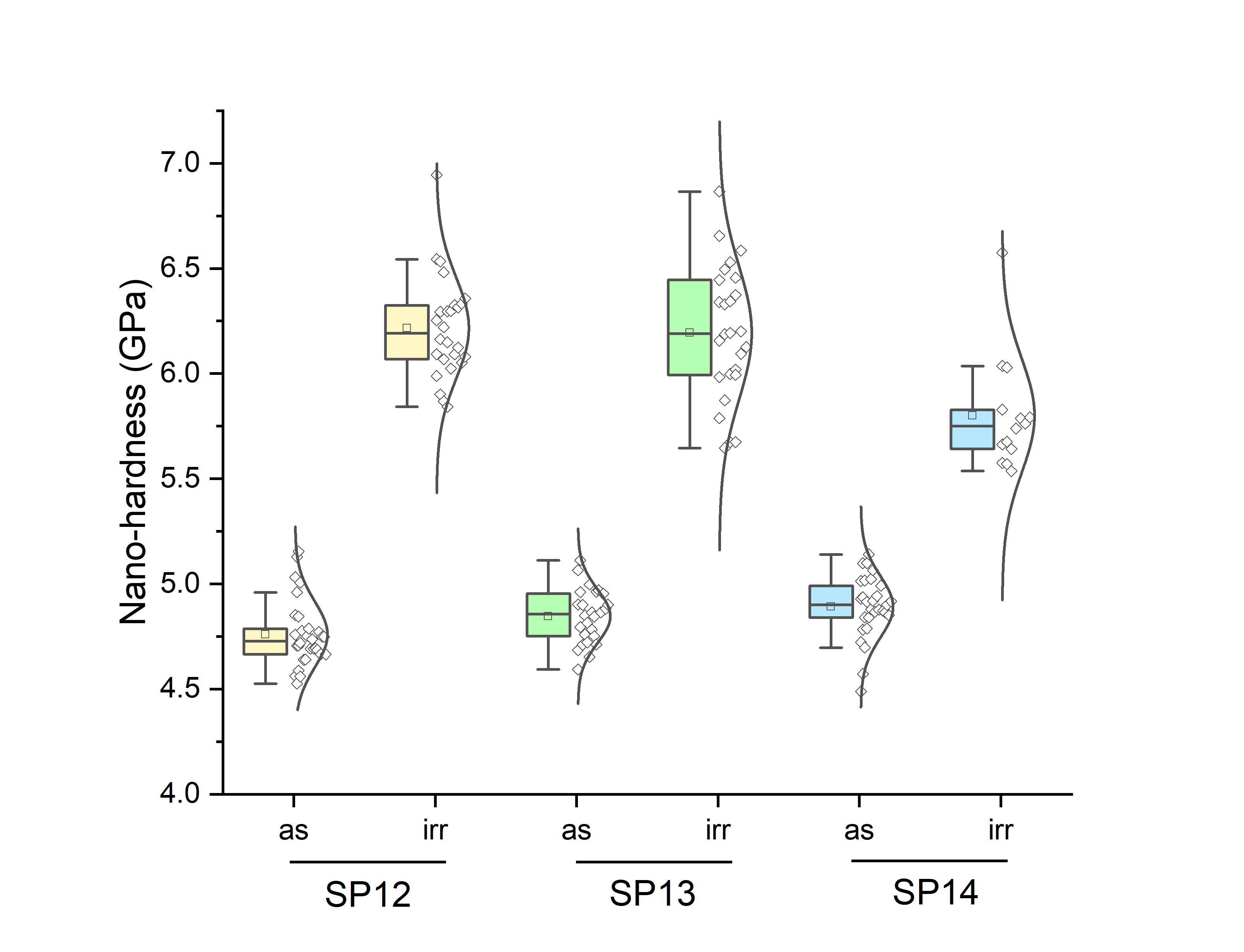}
\caption{The bulk-equivalent hardness calculated by Nix-Gao method with the data from 100nm to 1000nm of the unirradiated and from 100nm to 300nm of the irradiated materials.}\label{fig:ni}
\end{figure}

In this study, we chose the 100nm-300nm as the standard range to evaluate bulk-hardness for irradiated materials, because in this range, the indentation deforming zone is fully within the irradiated region. Fig.\ref{fig:ni} summarized the bulk-equivalent hardness of each nanoindentation of the irradiated and unirradiated steels. In unirradiated steels, the hardness linearly increases with Cr concentration. In irradiated materials, Fe12Cr9Al and Fe15Cr9Al showed similar high hardness while Fe18Cr9Al had the lowest.
 
\section{Discussion} 

\begin{figure}[btp] 
\centering
\includegraphics[width=\linewidth]{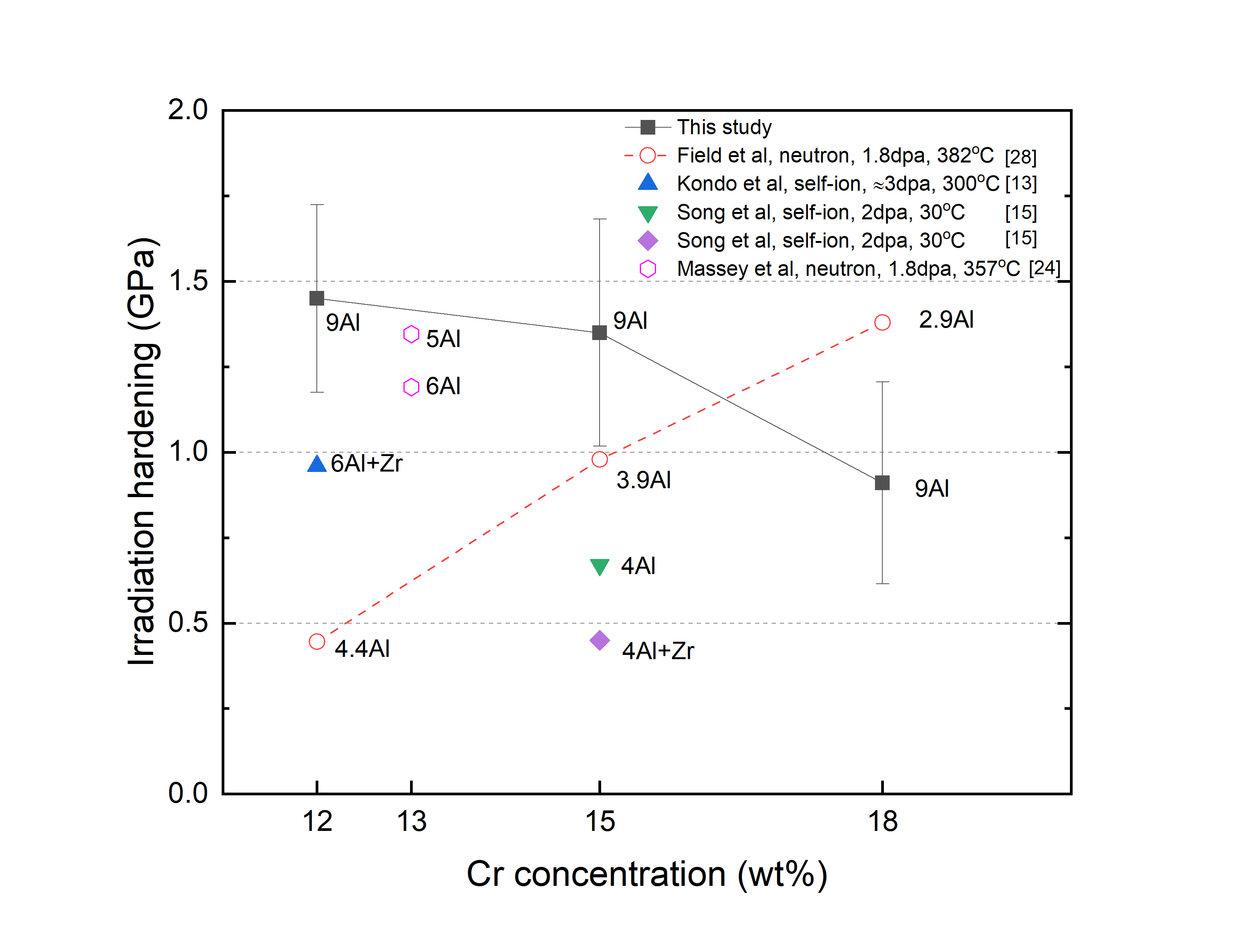}
\caption{The summary of Cr dependent irradiation hardening in FeCrAl steels. The solid symbols stand for ODS steels, and hollow symbols are non-ODS steels. The neutron irradiation hardening measured by yield stress were converted to nanoindentation hardening by NH(GPa)=0.004891YS(MPa).  }\label{fig:hardening}
\end{figure}
Figure \ref{fig:hardening} summarized the irradiation hardening in several FeCrAl ODS and non-ODS steels at similar irradiation conditions\citep{kondoIonIrradiationEffects2018}\citep{songInsightsHardeningPlastically2022}\citep{masseyPostIrradiationExamination2019}\citep{fieldPrecipitationNeutronIrradiated2018}. In this study, the irradiation hardening exhibited Cr dependence, where a higher Cr content yielded a smaller irradiation hardening. However, this trend is contradicted to the conventional understanding that high Cr should generate more $\alpha^\prime$ precipitate, thus yield a higher irradiation hardening. This trend is also opposite to the hardening of FeCrAl non-ODS steels which were subjected to neutron irradiation reported by Field et al\citep{fieldPrecipitationNeutronIrradiated2018} (Fig.\ref{fig:hardening}). There is a small successive decrease of the Al concentration in the studied steels (Table \ref{tab:1}), which might affect the formation of Al-enrich $\beta^\prime$ precipitates, and consistent with the decreasing trend of irradiation hardening. However, the effect is doubtful because of the tiny difference of the Al concentrations ($<1\%$). 
 
One explanation might be that the $\alpha^\prime$ was not precipitate at all under current irradiation flux. Ke et al\citep{keFluxEffectsPrecipitation2019} simulated the Cr precipitate in Fe-Cr binary system based on phase-field theory. The results showed that increasing dose rate will suppress the formation of $\alpha^\prime$ at 300\textdegree C. The threshold flux of $\alpha^\prime$ disappearing was as low as $10^{-4} dpa/s$. The experiments analysis by Zhao et al\citep{zhaoEffectHeavyIon2022} on the precipitates using APT demonstrated that $\alpha^\prime$ will dissolve under $10^{-3} dpa/s$ ion-irradiation at 300\textdegree C. Therefore, it is reasonable to conclude that $\alpha^\prime$ is not the main factor in irradiation hardening at the experiment condition in this study (dose rate $10^{-4}~10^{-3} dpa/s$). Oppositely, in the neutron irradiated FeCrAl steels, the main contributor of hardening was $\alpha^\prime$ precipitates under the dose rate of $8.1\times 10^{-7} dpa/s$, that the higher Cr content led to higher hardening\citep{fieldPrecipitationNeutronIrradiated2018}.

The hardening in FeCrAl ODS steels in this study was mainly owing to the dislocation loops. The solution atoms of Cr could play an essential role in the evolution of dislocation loops under irradiations\citep{maoImprovedIrradiationResistance2022}. Firstly, the solute element Cr could impede the growth of dislocation loops. This effect will make loops grow into small sizes. 

Secondly, the Cr will hinder the migration of $1/2<111>$ loops. The formation of $<100>$ loops by two glissile $1/2<111>$ reaction will be suppressed\citep{haleyDislocationLoopEvolution2017}, and the sessile $<100>$ loops are considered as stronger obstacles for dislocation migration than $1/2<111>$ loops. Thirdly, the formation energy of $<100>$ loops is much larger than $1/2<111>$ loops, and the energy increasing much faster with Cr concentration in $<100>$ loops than the latter\citep{zhangEffectCrConcentration2021}. Thus, the irradiation hardening will be reduced by suppressing formation of dislocation loops with elevated Cr content. 

Cottrell et al\citep{cottrellImmobilizationInterstitialLoops2004} developed a method to evalutate the migration activation energy of dislocation loops with respect to element content in steels. Here we simplified this method to one equation as:

\begin{equation}\label{eq:2}
E_{M}\left( n \right)\approx 0.92n^{1/3}\cdot U_{0}^{8/9}\cdot c^{4/9}
\end{equation}

Where $n$ is the number of atomic spacing in a dislocation segment, $U_0$ is the binding energy of atom in the core of edge dislocation, c is the concentration of alloying element. In the case of Cr, $U_0$=0.032 eV.

For the typical value n=25, the activation energy by Cr atoms is 0.048 eV for Fe12Cr9Al, 0.053 eV for Fe15Cr9Al, and 0.056 eV for Fe18Cr9Al ODS steels, respectively. The activation energy is increased with the Cr content; however, the increases are very small in comparison to Al (0.764 eV). 
 
The examination of the dislocation loops was performed in Fe12Cr9Al and Fe15Cr9Al ODS steels. As the samples are limited, only the information of total dislocation loops was compared (Table \ref{tab:loops}). Generally, the dislocation loops in Fe12Cr9Al are slightly smaller and denser than in Fe15Cr9Cr ODS steels, which is partially consistent with the above analysis.
 
The simplified dispersed barrier hardening (DBH) model (equation \ref{eq:3}) is used to evaluate the hardening induced by dislocation loops. In equation \ref{eq:3}, $M=3.06$ is the Taylor factor for FCC and BCC materials. The coefficient $\alpha =0.2$ is a strength factor. The shear modulus $\mu$ is 82 GPa for FeCrAl ODS steels. The Burgers vector b is 0.249 nm for glissile $1/2<111>$ dislocations. $N$ and $d$ present for the number density and diameter of loops, respectively.

\begin{equation}\label{eq:3}
\Delta\sigma_{loop}=M\alpha\mu b\sqrt{Nd}
\end{equation}

The estimated hardening by dislocation loops for SP12 and SP13 are 1.28 GPa and 1.18 GPa respectively. The measured hardening by nanoindentation were $1.45\pm 0.27$ GPa and $1.35\pm 0.33$ GPa for SP12 and SP13 (Table \ref{tab:loops}). 

Other factors which may influence the hardening are defect sink position including grain boundaries, pre-existed dislocations, and oxides. In the solid-solution strengthening analysis by Ukai et al\citep{ukaiSolidsolutionStrengtheningCr2021}, the grain sizes of the same SP-series steels are measured the same, and the dislocation densities are considered similar. Here we follow the same assumption. The oxides, however, may contribute to the differences. On one hand, the surface of oxides are effective annihilation sites for point defects such as interstitials and vacancies. On the other hand, atoms in large oxides may subject to knock-out effects, dissolute into matrix and reprecipitate as satellite clusters around original oxide\citep{songAssessmentPhaseStability2020a}. This progress might explain the decreasing trend of hardening, with the increased number density of oxides, which offered larger surface area in total, and decreased diameter, which reduced the possibility of dissolution (Fig.\ref{fig:fraction}).

Moreover, it seems reducing Al and adding Zr will reduce the irradiation hardening, according to the hardening results from Kondo et al\citep{kondoIonIrradiationEffects2018} and Song et al\citep{songInsightsHardeningPlastically2022} (Fig.\ref{fig:hardening}). Although there are some differences in the irradiation condition in Fe15Cr4Al(Zr), with lower temperature (30 \textdegree C) and damage dose (2 dpa), the results are reasonable by the explanation that the oxide density is much higher in steels with 4wt\%Al+Zr than with 9wt\%Al, where irradiation defects were annihilated by the larger oxide surface area.
 
Lower irradiation hardening usually indicates better irradiation resistance, though it is not absolutely correct due to various assessment standards. The irradiation hardening results in this study suggest the mechanism of Cr dependence in ion-irradiation is different to neutron irradiation. The current study in high Al FeCrAl steels will help to understand materials behavior at the irradiation conditions. It may contribute to the FeCrAl material design near the safe boundary of element concentration. 
 
\section{Conclusion}
The Cr dependent irradiation hardening in FeCrAl ODS steels with high Al concentration was studied. Three FeCrAl ODS steels, Fe12Cr9Al (SP12), Fe15Cr9Al (SP13), and Fe18Cr9Al (SP14), were irradiated with 6.4 MeV $Fe^{3+}$ at 300 \textdegree C to nominal 3 dpa. Both $1/2<111>$ and $<100>$ dislocation loops formed after irradiation. No void appeared in the irradiated region. Oxides remained crystalline after irradiation at experiment temperature. The irradiation hardening was investigated by nanoindentation and TEM. The hardening decreases with increasing Cr in 9w\% Al FeCrAl ODS steels. Particularly, the Fe18Cr9Al showed the lowest irradiation hardening in the experiment condition. Due to the high dose rate in heavy ion-irradiations, $\alpha^\prime$ should not be the main factor to irradiation hardening. The solute Cr atoms are considered to hinder the growth of dislocation loops formation, that reduced the irradiation hardening. Besides, oxides might also influence the hardening. The results will give reference for material design with high Al in FeCrAl ODS steels. 

\section{Acknowledgement}
We gratefully acknowledge Prof. Kondo and Mr. Hashitomi for their helping with the ion irradiation in DuET, Kyoto University. ZXZ acknowledges Prof. Stuart A. Maloy in Los Alamos National Laboratory for the discussion ATF element selection. ZXZ acknowledges Prof. Kasada in Tohoku University for discussion on nanoindentation. ZXZ would like to thank Prof. Akihiko Kimura in Kyoto University for the supporting of the experiment and the research communications.  




  \bibliographystyle{elsarticle-num-names} 
  \bibliography{My_Library.bib}

\begin{thebibliography}{47}
\expandafter\ifx\csname natexlab\endcsname\relax\def\natexlab#1{#1}\fi
\providecommand{\url}[1]{\texttt{#1}}
\providecommand{\href}[2]{#2}
\providecommand{\path}[1]{#1}
\providecommand{\DOIprefix}{doi:}
\providecommand{\ArXivprefix}{arXiv:}
\providecommand{\URLprefix}{URL: }
\providecommand{\Pubmedprefix}{pmid:}
\providecommand{\doi}[1]{\href{http://dx.doi.org/#1}{\path{#1}}}
\providecommand{\Pubmed}[1]{\href{pmid:#1}{\path{#1}}}
\providecommand{\bibinfo}[2]{#2}
\ifx\xfnm\relax \def\xfnm[#1]{\unskip,\space#1}\fi
\bibitem[{Zinkle et~al.(2014)Zinkle, Terrani, Gehin, Ott, and
  Snead}]{zinkleAccidentTolerantFuels2014}
\bibinfo{author}{S.~J. Zinkle}, \bibinfo{author}{K.~A. Terrani},
  \bibinfo{author}{J.~C. Gehin}, \bibinfo{author}{L.~J. Ott},
  \bibinfo{author}{L.~L. Snead},
\newblock \bibinfo{title}{Accident tolerant fuels for {{LWRs}}: {{A}}
  perspective},
\newblock \bibinfo{journal}{Journal of Nuclear Materials} \bibinfo{volume}{448}
  (\bibinfo{year}{2014}) \bibinfo{pages}{374--379}.
\bibitem[{Terrani(2018)}]{terraniAccidentTolerantFuel2018}
\bibinfo{author}{K.~A. Terrani},
\newblock \bibinfo{title}{Accident tolerant fuel cladding development:
  {{Promise}}, status, and challenges},
\newblock \bibinfo{journal}{Journal of Nuclear Materials} \bibinfo{volume}{501}
  (\bibinfo{year}{2018}) \bibinfo{pages}{13--30}.
\bibitem[{Kimura et~al.(2003)Kimura, Ukai, and
  Fujiwara}]{kimuraDevelopmentFuelClad2003}
\bibinfo{author}{A.~Kimura}, \bibinfo{author}{S.~Ukai},
  \bibinfo{author}{M.~Fujiwara},
\newblock \bibinfo{title}{Development of fuel clad materials for high burn-up
  operation of {{SCPR}}},
\newblock \bibinfo{journal}{Proc. GENES4/ANP2003, Paper} \bibinfo{volume}{1198}
  (\bibinfo{year}{2003}).
\bibitem[{Kimura et~al.(2009)Kimura, Kasada, Iwata, Kishimoto, Zhang, Isselin,
  Dou, Lee, Muthukumar, and Okuda}]{kimuraSuperODSSteels2009}
\bibinfo{author}{A.~Kimura}, \bibinfo{author}{R.~Kasada},
  \bibinfo{author}{N.~Iwata}, \bibinfo{author}{H.~Kishimoto},
  \bibinfo{author}{{\relax CH}.~Zhang}, \bibinfo{author}{J.~Isselin},
  \bibinfo{author}{P.~Dou}, \bibinfo{author}{{\relax JH}.~Lee},
  \bibinfo{author}{N.~Muthukumar}, \bibinfo{author}{T.~Okuda},
\newblock \bibinfo{title}{Super {{ODS}} steels {{R}}\&{{D}} for fuel cladding
  of next generation nuclear systems 1) {{Introduction}} and alloy design},
\newblock in: \bibinfo{booktitle}{International {{Congress}} on {{Advances}} in
  {{Nuclear Power Plants}} 2009, {{ICAPP}} 2009}, \bibinfo{publisher}{{Atomic
  Energy Society of Japan}}, \bibinfo{year}{2009}, pp.
  \bibinfo{pages}{2187--2194}.
\bibitem[{Kimura et~al.(2011)Kimura, Kasada, Iwata, Kishimoto, Zhang, Isselin,
  Dou, Lee, Muthukumar, and Okuda}]{kimuraDevelopmentAddedHighCr2011}
\bibinfo{author}{A.~Kimura}, \bibinfo{author}{R.~Kasada},
  \bibinfo{author}{N.~Iwata}, \bibinfo{author}{H.~Kishimoto},
  \bibinfo{author}{{\relax CH}.~Zhang}, \bibinfo{author}{J.~Isselin},
  \bibinfo{author}{P.~Dou}, \bibinfo{author}{{\relax JH}.~Lee},
  \bibinfo{author}{N.~Muthukumar}, \bibinfo{author}{T.~Okuda},
\newblock \bibinfo{title}{Development of {{Al}} added high-{{Cr ODS}} steels
  for fuel cladding of next generation nuclear systems},
\newblock \bibinfo{journal}{Journal of Nuclear Materials} \bibinfo{volume}{417}
  (\bibinfo{year}{2011}) \bibinfo{pages}{176--179}.
\bibitem[{Maloy et~al.(2016{\natexlab{a}})Maloy, Aydogan, Anderoglu, Lavender,
  and Yamamoto}]{maloyViabilityThinWall2016}
\bibinfo{author}{{\relax SA}.~Maloy}, \bibinfo{author}{E.~Aydogan},
  \bibinfo{author}{O.~Anderoglu}, \bibinfo{author}{C.~Lavender},
  \bibinfo{author}{Y.~Yamamoto},
\newblock \bibinfo{title}{Viability of thin wall tube forming of {{ATF
  FeCrAl}}.; {{Los Alamos National Lab}}.({{LANL}}), {{Los Alamos}}, {{NM}}
  ({{United States}})},
\newblock \bibinfo{journal}{Pacific Northwest National Lab.(PNNL), Richland, WA
  (United States)}  (\bibinfo{year}{2016}{\natexlab{a}}).
\bibitem[{Maloy et~al.(2016{\natexlab{b}})Maloy, Saleh, Anderoglu, Romero,
  Odette, Yamamoto, Li, Cole, and
  Fielding}]{maloyCharacterizationComparativeAnalysis2016}
\bibinfo{author}{S.~A. Maloy}, \bibinfo{author}{T.~A. Saleh},
  \bibinfo{author}{O.~Anderoglu}, \bibinfo{author}{T.~J. Romero},
  \bibinfo{author}{G.~R. Odette}, \bibinfo{author}{T.~Yamamoto},
  \bibinfo{author}{S.~Li}, \bibinfo{author}{J.~I. Cole},
  \bibinfo{author}{R.~Fielding},
\newblock \bibinfo{title}{Characterization and comparative analysis of the
  tensile properties of five tempered martensitic steels and an oxide
  dispersion strengthened ferritic alloy irradiated at
  {$\approx$}295~\textdegree{{C}} to {$\approx$}6.5~dpa},
\newblock \bibinfo{journal}{Journal of Nuclear Materials} \bibinfo{volume}{468}
  (\bibinfo{year}{2016}{\natexlab{b}}) \bibinfo{pages}{232--239}.
  \DOIprefix\doi{10.1016/j.jnucmat.2015.07.039}.
\bibitem[{Kobayashi and Takasugi(2010)}]{kobayashiMapping475Embrittlement2010}
\bibinfo{author}{S.~Kobayashi}, \bibinfo{author}{T.~Takasugi},
\newblock \bibinfo{title}{Mapping of 475 \textdegree{{C}} embrittlement in
  ferritic {{Fe}}\textendash{{Cr}}\textendash{{Al}} alloys},
\newblock \bibinfo{journal}{Scripta Materialia} \bibinfo{volume}{63}
  (\bibinfo{year}{2010}) \bibinfo{pages}{1104--1107}.
  \DOIprefix\doi{10.1016/j.scriptamat.2010.08.015}.
\bibitem[{Chen et~al.(2015)Chen, Kimura, Han, and
  Je}]{chenAgehardeningSusceptibilityHighCr2015}
\bibinfo{author}{D.~Chen}, \bibinfo{author}{A.~Kimura},
  \bibinfo{author}{W.~Han}, \bibinfo{author}{H.~Je},
\newblock \bibinfo{title}{Age-hardening susceptibility of high-{{Cr ODS}}
  ferritic steels and {{SUS430}} ferritic steel},
\newblock \bibinfo{journal}{Fusion Engineering and Design}
  \bibinfo{volume}{98--99} (\bibinfo{year}{2015}) \bibinfo{pages}{1945--1949}.
  \DOIprefix\doi{10.1016/j.fusengdes.2015.05.078}.
\bibitem[{Maji et~al.(2021)Maji, Ukai, and
  {Oono-Hori}}]{majiMicrostructuralStabilityIntermetallic2021}
\bibinfo{author}{B.~C. Maji}, \bibinfo{author}{S.~Ukai},
  \bibinfo{author}{N.~{Oono-Hori}},
\newblock \bibinfo{title}{Microstructural stability and intermetallic
  embrittlement in high {{Al}} containing {{FeCrAl-ODS}} alloys},
\newblock \bibinfo{journal}{Materials Science and Engineering: A}
  \bibinfo{volume}{807} (\bibinfo{year}{2021}) \bibinfo{pages}{140858}.
  \DOIprefix\doi{10.1016/j.msea.2021.140858}.
\bibitem[{Sang et~al.(2020)Sang, Dou, and
  Kimura}]{sangEarlystageThermalAgeing2020}
\bibinfo{author}{W.~Sang}, \bibinfo{author}{P.~Dou},
  \bibinfo{author}{A.~Kimura},
\newblock \bibinfo{title}{Early-stage thermal ageing behavior of {{12Cr}},
  {{12Cr}}\textendash{{7Al}} and {{18Cr}}\textendash{{9Al ODS}} steels},
\newblock \bibinfo{journal}{Journal of Nuclear Materials} \bibinfo{volume}{535}
  (\bibinfo{year}{2020}) \bibinfo{pages}{152164}.
  \DOIprefix\doi{10.1016/j.jnucmat.2020.152164}.
\bibitem[{Kimura et~al.(2023)Kimura, Sang, Han, Yabuuchi, Xin, Luan, and
  Dou}]{kimuraTwofoldAgehardeningMechanism2023}
\bibinfo{author}{A.~Kimura}, \bibinfo{author}{W.~Sang},
  \bibinfo{author}{W.~Han}, \bibinfo{author}{K.~Yabuuchi},
  \bibinfo{author}{Z.~Xin}, \bibinfo{author}{J.~Luan},
  \bibinfo{author}{P.~Dou},
\newblock \bibinfo{title}{Twofold age-hardening mechanism of {{Al-added}}
  high-{{Cr ODS}} ferritic steels},
\newblock \bibinfo{journal}{Journal of Nuclear Materials} \bibinfo{volume}{575}
  (\bibinfo{year}{2023}) \bibinfo{pages}{154223}.
\bibitem[{Kondo et~al.(2018)Kondo, Aoki, Yamashita, Ukai, Sakamoto, Hirai, and
  Kimura}]{kondoIonIrradiationEffects2018}
\bibinfo{author}{K.~Kondo}, \bibinfo{author}{S.~Aoki},
  \bibinfo{author}{S.~Yamashita}, \bibinfo{author}{S.~Ukai},
  \bibinfo{author}{K.~Sakamoto}, \bibinfo{author}{M.~Hirai},
  \bibinfo{author}{A.~Kimura},
\newblock \bibinfo{title}{Ion irradiation effects on {{FeCrAl-ODS}} ferritic
  steel},
\newblock \bibinfo{journal}{Nuclear Materials and Energy} \bibinfo{volume}{15}
  (\bibinfo{year}{2018}) \bibinfo{pages}{13--16}.
  \DOIprefix\doi{10.1016/j.nme.2018.05.022}.
\bibitem[{Song et~al.(2018)Song, Morrall, Zhang, Yabuuchi, and
  Kimura}]{songRadiationResponseODS2018}
\bibinfo{author}{P.~Song}, \bibinfo{author}{D.~Morrall},
  \bibinfo{author}{Z.~Zhang}, \bibinfo{author}{K.~Yabuuchi},
  \bibinfo{author}{A.~Kimura},
\newblock \bibinfo{title}{Radiation response of {{ODS}} ferritic steels with
  different oxide particles under ion-irradiation at 550 {{C}}},
\newblock \bibinfo{journal}{Journal of Nuclear Materials} \bibinfo{volume}{502}
  (\bibinfo{year}{2018}) \bibinfo{pages}{76--85}.
\bibitem[{Song et~al.(2022)Song, Yabuuchi, and
  Sp{\"a}tig}]{songInsightsHardeningPlastically2022}
\bibinfo{author}{P.~Song}, \bibinfo{author}{K.~Yabuuchi},
  \bibinfo{author}{P.~Sp{\"a}tig},
\newblock \bibinfo{title}{Insights into hardening, plastically deformed zone
  and geometrically necessary dislocations of two ion-irradiated {{FeCrAl}}
  ({{Zr}})-{{ODS}} ferritic steels: {{A}} combined experimental and simulation
  study},
\newblock \bibinfo{journal}{Acta Materialia} \bibinfo{volume}{234}
  (\bibinfo{year}{2022}) \bibinfo{pages}{117991}.
\bibitem[{Gao et~al.(2020)Gao, Yamasaki, Song, Huang, Yabuuchi, Kimura,
  Sakamoto, and Yamashita}]{gaoDoseDependenceIon2020}
\bibinfo{author}{J.~Gao}, \bibinfo{author}{Y.~Yamasaki},
  \bibinfo{author}{P.~Song}, \bibinfo{author}{Y.-J. Huang},
  \bibinfo{author}{K.~Yabuuchi}, \bibinfo{author}{A.~Kimura},
  \bibinfo{author}{K.~Sakamoto}, \bibinfo{author}{S.~Yamashita},
\newblock \bibinfo{title}{Dose dependence of ion irradiation effects on
  {{12Cr}}\textendash{{6Al-ODS}} steel with electron-beam weld line},
\newblock \bibinfo{journal}{Journal of Nuclear Materials} \bibinfo{volume}{528}
  (\bibinfo{year}{2020}) \bibinfo{pages}{151858}.
\bibitem[{Getto et~al.(2022{\natexlab{a}})Getto, Johnson, Maughan, Nathan,
  McMahan, Baker, Knipling, Briggs, Hattar, and
  Swenson}]{gettoFrictionStirWelding2022}
\bibinfo{author}{E.~Getto}, \bibinfo{author}{M.~Johnson},
  \bibinfo{author}{M.~Maughan}, \bibinfo{author}{N.~Nathan},
  \bibinfo{author}{J.~McMahan}, \bibinfo{author}{B.~Baker},
  \bibinfo{author}{K.~Knipling}, \bibinfo{author}{S.~Briggs},
  \bibinfo{author}{K.~Hattar}, \bibinfo{author}{M.~J. Swenson},
\newblock \bibinfo{title}{Friction stir welding and self-ion irradiation
  effects on microstructure and mechanical properties changes within oxide
  dispersion strengthened steel {{MA956}}},
\newblock \bibinfo{journal}{Journal of Nuclear Materials} \bibinfo{volume}{567}
  (\bibinfo{year}{2022}{\natexlab{a}}) \bibinfo{pages}{153795}.
  \DOIprefix\doi{10.1016/j.jnucmat.2022.153795}.
\bibitem[{Getto et~al.(2022{\natexlab{b}})Getto, Nathan, McMahan, Taller, and
  Baker}]{gettoUnderstandingRadiationEffects2022}
\bibinfo{author}{E.~Getto}, \bibinfo{author}{N.~Nathan},
  \bibinfo{author}{J.~McMahan}, \bibinfo{author}{S.~Taller},
  \bibinfo{author}{B.~Baker},
\newblock \bibinfo{title}{Understanding radiation effects in friction stir
  welded {{MA956}} using ion irradiation and a rate theory model},
\newblock \bibinfo{journal}{Journal of Nuclear Materials} \bibinfo{volume}{561}
  (\bibinfo{year}{2022}{\natexlab{b}}) \bibinfo{pages}{153530}.
  \DOIprefix\doi{10.1016/j.jnucmat.2022.153530}.
\bibitem[{Ejenstam et~al.(2015)Ejenstam, Thuvander, Olsson, Rave, and
  Szakalos}]{ejenstamMicrostructuralStabilityFe2015}
\bibinfo{author}{J.~Ejenstam}, \bibinfo{author}{M.~Thuvander},
  \bibinfo{author}{P.~Olsson}, \bibinfo{author}{F.~Rave},
  \bibinfo{author}{P.~Szakalos},
\newblock \bibinfo{title}{Microstructural stability of
  {{Fe}}\textendash{{Cr}}\textendash{{Al}} alloys at 450\textendash 550
  \textdegree{{C}}},
\newblock \bibinfo{journal}{Journal of Nuclear Materials} \bibinfo{volume}{457}
  (\bibinfo{year}{2015}) \bibinfo{pages}{291--297}.
  \DOIprefix\doi{10.1016/j.jnucmat.2014.11.101}.
\bibitem[{Zhou et~al.(2019)Zhou, Guo, Wei, Wang, Chen, Chen, Zhang, Liu, Liu,
  and Mo}]{zhouEffectAluminumContent2019}
\bibinfo{author}{X.~Zhou}, \bibinfo{author}{L.~Guo}, \bibinfo{author}{Y.~Wei},
  \bibinfo{author}{H.~Wang}, \bibinfo{author}{C.~Chen},
  \bibinfo{author}{Y.~Chen}, \bibinfo{author}{W.~Zhang},
  \bibinfo{author}{S.~Liu}, \bibinfo{author}{R.~Liu}, \bibinfo{author}{S.~Mo},
\newblock \bibinfo{title}{Effect of aluminum content on dislocation loops in
  model {{FeCrAl}} alloys},
\newblock \bibinfo{journal}{Nuclear Materials and Energy} \bibinfo{volume}{21}
  (\bibinfo{year}{2019}) \bibinfo{pages}{100718}.
\bibitem[{Reese et~al.(2018)Reese, Almirall, Yamamoto, Tumey, Robert~Odette,
  and Marquis}]{reeseDoseRateDependence2018}
\bibinfo{author}{E.~R. Reese}, \bibinfo{author}{N.~Almirall},
  \bibinfo{author}{T.~Yamamoto}, \bibinfo{author}{S.~Tumey},
  \bibinfo{author}{G.~Robert~Odette}, \bibinfo{author}{E.~A. Marquis},
\newblock \bibinfo{title}{Dose rate dependence of {{Cr}} precipitation in an
  ion-irradiated {{Fe18Cr}} alloy},
\newblock \bibinfo{journal}{Scripta Materialia} \bibinfo{volume}{146}
  (\bibinfo{year}{2018}) \bibinfo{pages}{213--217}.
  \DOIprefix\doi{10.1016/j.scriptamat.2017.11.040}.
\bibitem[{Edmondson et~al.(2016)Edmondson, Briggs, {Y.Yamamoto}, Howard,
  Sridharan, Terrani, and
  Field}]{edmondsonIrradiationenhancedPrecipitationModel2016}
\bibinfo{author}{P.~Edmondson}, \bibinfo{author}{S.~Briggs},
  \bibinfo{author}{{Y.Yamamoto}}, \bibinfo{author}{R.~Howard},
  \bibinfo{author}{K.~Sridharan}, \bibinfo{author}{K.~Terrani},
  \bibinfo{author}{K.~Field},
\newblock \bibinfo{title}{Irradiation-enhanced {$\alpha{'}$} precipitation in
  model {{FeCrAl}} alloys},
\newblock \bibinfo{journal}{Scripta Materialia} \bibinfo{volume}{116}
  (\bibinfo{year}{2016}) \bibinfo{pages}{112--116}.
  \DOIprefix\doi{10.1016/j.scriptamat.2016.02.002}.
\bibitem[{Massey et~al.(2019)Massey, Edmondson, Field, Hoelzer, Dryepondt,
  Terrani, and Zinkle}]{masseyPostIrradiationExamination2019}
\bibinfo{author}{C.~P. Massey}, \bibinfo{author}{P.~D. Edmondson},
  \bibinfo{author}{K.~G. Field}, \bibinfo{author}{D.~T. Hoelzer},
  \bibinfo{author}{S.~N. Dryepondt}, \bibinfo{author}{K.~A. Terrani},
  \bibinfo{author}{S.~J. Zinkle},
\newblock \bibinfo{title}{Post irradiation examination of nanoprecipitate
  stability and {$\alpha{'}$} precipitation in an oxide dispersion strengthened
  {{Fe-12Cr-5Al}} alloy},
\newblock \bibinfo{journal}{Scripta Materialia} \bibinfo{volume}{162}
  (\bibinfo{year}{2019}) \bibinfo{pages}{94--98}.
\bibitem[{Zhang et~al.(2019)Zhang, Briggs, Edmondson, Gussev, Howard, and
  Field}]{zhangInfluenceWeldingNeutron2019}
\bibinfo{author}{D.~Zhang}, \bibinfo{author}{S.~A. Briggs},
  \bibinfo{author}{P.~D. Edmondson}, \bibinfo{author}{M.~N. Gussev},
  \bibinfo{author}{R.~H. Howard}, \bibinfo{author}{K.~G. Field},
\newblock \bibinfo{title}{Influence of welding and neutron irradiation on
  dislocation loop formation and {$\alpha{'}$} precipitation in a {{FeCrAl}}
  alloy},
\newblock \bibinfo{journal}{Journal of Nuclear Materials} \bibinfo{volume}{527}
  (\bibinfo{year}{2019}) \bibinfo{pages}{151784}.
\bibitem[{Mao et~al.(2022)Mao, Massey, Yamamoto, Unocic, Gussev, Zhang, Briggs,
  Karakoc, Nelson, and Field}]{maoImprovedIrradiationResistance2022}
\bibinfo{author}{K.~S. Mao}, \bibinfo{author}{C.~P. Massey},
  \bibinfo{author}{Y.~Yamamoto}, \bibinfo{author}{K.~A. Unocic},
  \bibinfo{author}{M.~N. Gussev}, \bibinfo{author}{D.~Zhang},
  \bibinfo{author}{S.~A. Briggs}, \bibinfo{author}{O.~Karakoc},
  \bibinfo{author}{A.~T. Nelson}, \bibinfo{author}{K.~G. Field},
\newblock \bibinfo{title}{Improved irradiation resistance of accident-tolerant
  high-strength {{FeCrAl}} alloys with heterogeneous structures},
\newblock \bibinfo{journal}{Acta Materialia} \bibinfo{volume}{231}
  (\bibinfo{year}{2022}) \bibinfo{pages}{117843}.
\bibitem[{Field et~al.(2018)Field, Littrell, and
  Briggs}]{fieldPrecipitationNeutronIrradiated2018}
\bibinfo{author}{K.~G. Field}, \bibinfo{author}{K.~C. Littrell},
  \bibinfo{author}{S.~A. Briggs},
\newblock \bibinfo{title}{Precipitation of {$\alpha{'}$} in neutron irradiated
  commercial {{FeCrAl}} alloys},
\newblock \bibinfo{journal}{Scripta Materialia} \bibinfo{volume}{142}
  (\bibinfo{year}{2018}) \bibinfo{pages}{41--45}.
\bibitem[{Field et~al.(2015)Field, Hu, Littrell, Yamamoto, and
  Snead}]{fieldRadiationToleranceNeutronirradiated2015}
\bibinfo{author}{K.~G. Field}, \bibinfo{author}{X.~Hu}, \bibinfo{author}{K.~C.
  Littrell}, \bibinfo{author}{Y.~Yamamoto}, \bibinfo{author}{L.~L. Snead},
\newblock \bibinfo{title}{Radiation tolerance of neutron-irradiated model
  {{Fe}}\textendash{{Cr}}\textendash{{Al}} alloys},
\newblock \bibinfo{journal}{Journal of Nuclear Materials} \bibinfo{volume}{465}
  (\bibinfo{year}{2015}) \bibinfo{pages}{746--755}.
\bibitem[{Aydogan et~al.(2018)Aydogan, Weaver, Maloy, {El-Atwani}, Wang, and
  Mara}]{aydoganMicrostructureMechanicalProperties2018}
\bibinfo{author}{E.~Aydogan}, \bibinfo{author}{J.~S. Weaver},
  \bibinfo{author}{S.~A. Maloy}, \bibinfo{author}{O.~{El-Atwani}},
  \bibinfo{author}{Y.~Q. Wang}, \bibinfo{author}{N.~A. Mara},
\newblock \bibinfo{title}{Microstructure and mechanical properties of
  {{FeCrAl}} alloys under heavy ion irradiations},
\newblock \bibinfo{journal}{Journal of Nuclear Materials} \bibinfo{volume}{503}
  (\bibinfo{year}{2018}) \bibinfo{pages}{250--262}.
\bibitem[{Ukai et~al.(2023)Ukai, Sakamoto, Ohtsuka, Yamashita, and
  Kimura}]{ukaiAlloyDesignCharacterization2023}
\bibinfo{author}{S.~Ukai}, \bibinfo{author}{K.~Sakamoto},
  \bibinfo{author}{S.~Ohtsuka}, \bibinfo{author}{S.~Yamashita},
  \bibinfo{author}{A.~Kimura},
\newblock \bibinfo{title}{Alloy {{Design}} and {{Characterization}} of a
  {{Recrystallized FeCrAl-ODS Cladding}} for {{Accident-Tolerant BWR Fuels}}:
  {{An Overview}} of {{Research Activity}} in {{Japan}}},
\newblock \bibinfo{journal}{Journal of Nuclear Materials}
  (\bibinfo{year}{2023}) \bibinfo{pages}{154508}.
\bibitem[{Kohyama et~al.(2000)Kohyama, Katoh, Ando, and
  Jimbo}]{kohyamaNewMultipleBeams2000}
\bibinfo{author}{A.~Kohyama}, \bibinfo{author}{Y.~Katoh},
  \bibinfo{author}{M.~Ando}, \bibinfo{author}{K.~Jimbo},
\newblock \bibinfo{title}{A new {{Multiple Beams}}\textendash{{Material
  Interaction Research Facility}} for radiation damage studies in fusion
  materials},
\newblock \bibinfo{journal}{Fusion engineering and design} \bibinfo{volume}{51}
  (\bibinfo{year}{2000}) \bibinfo{pages}{789--795}.
\bibitem[{Ziegler et~al.(2010)Ziegler, Ziegler, and
  Biersack}]{zieglerSRIMStoppingRange2010}
\bibinfo{author}{J.~F. Ziegler}, \bibinfo{author}{M.~D. Ziegler},
  \bibinfo{author}{J.~P. Biersack},
\newblock \bibinfo{title}{{{SRIM}}\textendash{{The}} stopping and range of ions
  in matter (2010)},
\newblock \bibinfo{journal}{Nuclear Instruments and Methods in Physics Research
  Section B: Beam Interactions with Materials and Atoms} \bibinfo{volume}{268}
  (\bibinfo{year}{2010}) \bibinfo{pages}{1818--1823}.
\bibitem[{{ASTM E521-96 e1}(2009)}]{astme521-96e1StandardPracticeNeutron2009}
\bibinfo{author}{{ASTM E521-96 e1}},
\newblock \bibinfo{title}{Standard {{Practice}} for {{Neutron Radiation Damage
  Simulation}} by {{Charged Particle Radiation}}}  (\bibinfo{year}{2009}).
\bibitem[{Lucas and Oliver(1999)}]{lucasIndentationPowerlawCreep1999}
\bibinfo{author}{B.~N. Lucas}, \bibinfo{author}{W.~C. Oliver},
\newblock \bibinfo{title}{Indentation power-law creep of high-purity indium},
\newblock \bibinfo{journal}{Metallurgical and materials Transactions A}
  \bibinfo{volume}{30} (\bibinfo{year}{1999}) \bibinfo{pages}{601--610}.
\bibitem[{Oono et~al.(2017)Oono, Ukai, Hayashi, Ohtsuka, Kaito, Kimura,
  Torimaru, and Sakamoto}]{oonoGrowthOxideParticles2017}
\bibinfo{author}{N.~Oono}, \bibinfo{author}{S.~Ukai},
  \bibinfo{author}{S.~Hayashi}, \bibinfo{author}{S.~Ohtsuka},
  \bibinfo{author}{T.~Kaito}, \bibinfo{author}{A.~Kimura},
  \bibinfo{author}{T.~Torimaru}, \bibinfo{author}{K.~Sakamoto},
\newblock \bibinfo{title}{Growth of oxide particles in {{FeCrAl-}} oxide
  dispersion strengthened steels at high temperature},
\newblock \bibinfo{journal}{Journal of Nuclear Materials} \bibinfo{volume}{493}
  (\bibinfo{year}{2017}) \bibinfo{pages}{180--188}.
  \DOIprefix\doi{10.1016/j.jnucmat.2017.06.018}.
\bibitem[{Nix and Gao(1998)}]{nixIndentationSizeEffects1998}
\bibinfo{author}{W.~D. Nix}, \bibinfo{author}{H.~Gao},
\newblock \bibinfo{title}{Indentation size effects in crystalline materials: A
  law for strain gradient plasticity},
\newblock \bibinfo{journal}{Journal of the Mechanics and Physics of Solids}
  \bibinfo{volume}{46} (\bibinfo{year}{1998}) \bibinfo{pages}{411--425}.
\bibitem[{Yabuuchi et~al.(2014)Yabuuchi, Kuribayashi, Nogami, Kasada, and
  Hasegawa}]{yabuuchiEvaluationIrradiationHardening2014}
\bibinfo{author}{K.~Yabuuchi}, \bibinfo{author}{Y.~Kuribayashi},
  \bibinfo{author}{S.~Nogami}, \bibinfo{author}{R.~Kasada},
  \bibinfo{author}{A.~Hasegawa},
\newblock \bibinfo{title}{Evaluation of irradiation hardening of proton
  irradiated stainless steels by nanoindentation},
\newblock \bibinfo{journal}{Journal of nuclear materials} \bibinfo{volume}{446}
  (\bibinfo{year}{2014}) \bibinfo{pages}{142--147}.
\bibitem[{Busby et~al.(2005)Busby, Hash, and
  Was}]{busbyRelationshipHardnessYield2005}
\bibinfo{author}{J.~T. Busby}, \bibinfo{author}{M.~C. Hash},
  \bibinfo{author}{G.~S. Was},
\newblock \bibinfo{title}{The relationship between hardness and yield stress in
  irradiated austenitic and ferritic steels},
\newblock \bibinfo{journal}{Journal of Nuclear Materials} \bibinfo{volume}{336}
  (\bibinfo{year}{2005}) \bibinfo{pages}{267--278}.
\bibitem[{Song et~al.(2020)Song, Kimura, Yabuuchi, Dou, Watanabe, Gao, and
  Huang}]{songAssessmentPhaseStability2020a}
\bibinfo{author}{P.~Song}, \bibinfo{author}{A.~Kimura},
  \bibinfo{author}{K.~Yabuuchi}, \bibinfo{author}{P.~Dou},
  \bibinfo{author}{H.~Watanabe}, \bibinfo{author}{J.~Gao},
  \bibinfo{author}{Y.-J. Huang},
\newblock \bibinfo{title}{Assessment of phase stability of oxide particles in
  different types of {{15Cr-ODS}} ferritic steels under 6.4 {{MeV Fe}} ion
  irradiation at 200\textdegree{} {{C}}},
\newblock \bibinfo{journal}{Journal of Nuclear Materials} \bibinfo{volume}{529}
  (\bibinfo{year}{2020}) \bibinfo{pages}{151953}.
\bibitem[{Hau{\v s}ild(2021)}]{hausildMethodologicalCommentNanoindentation2021}
\bibinfo{author}{P.~Hau{\v s}ild},
\newblock \bibinfo{title}{Methodological comment on the nanoindentation of
  ion-irradiation hardened materials},
\newblock \bibinfo{journal}{Journal of Nuclear Materials} \bibinfo{volume}{551}
  (\bibinfo{year}{2021}) \bibinfo{pages}{152987}.
\bibitem[{Walter and Mitterer(2009)}]{walter3D2DFinite2009}
\bibinfo{author}{C.~Walter}, \bibinfo{author}{C.~Mitterer},
\newblock \bibinfo{title}{{{3D}} versus {{2D}} finite element simulation of the
  effect of surface roughness on nanoindentation of hard coatings},
\newblock \bibinfo{journal}{Surface and Coatings Technology}
  \bibinfo{volume}{203} (\bibinfo{year}{2009}) \bibinfo{pages}{3286--3290}.
\bibitem[{Zhu et~al.(2022)Zhu, Zhao, Agarwal, Henry, and
  Zinkle}]{zhuAccurateEvaluationBulk2022}
\bibinfo{author}{P.~Zhu}, \bibinfo{author}{Y.~Zhao},
  \bibinfo{author}{S.~Agarwal}, \bibinfo{author}{J.~Henry},
  \bibinfo{author}{S.~J. Zinkle},
\newblock \bibinfo{title}{Toward accurate evaluation of bulk hardness from
  nanoindentation testing at low indent depths},
\newblock \bibinfo{journal}{Materials \& Design} \bibinfo{volume}{213}
  (\bibinfo{year}{2022}) \bibinfo{pages}{110317}.
\bibitem[{Ke et~al.(2019)Ke, Reese, Marquis, Odette, and
  Morgan}]{keFluxEffectsPrecipitation2019}
\bibinfo{author}{J.-H. Ke}, \bibinfo{author}{E.~R. Reese},
  \bibinfo{author}{E.~A. Marquis}, \bibinfo{author}{G.~R. Odette},
  \bibinfo{author}{D.~Morgan},
\newblock \bibinfo{title}{Flux effects in precipitation under irradiation
  \textendash{} {{Simulation}} of {{Fe-Cr}} alloys},
\newblock \bibinfo{journal}{Acta Materialia} \bibinfo{volume}{164}
  (\bibinfo{year}{2019}) \bibinfo{pages}{586--601}.
  \DOIprefix\doi{10.1016/j.actamat.2018.10.063}.
\bibitem[{Zhao et~al.(2022)Zhao, Bhattacharya, Pareige, Massey, Zhu, Poplawsky,
  Henry, and Zinkle}]{zhaoEffectHeavyIon2022}
\bibinfo{author}{Y.~Zhao}, \bibinfo{author}{A.~Bhattacharya},
  \bibinfo{author}{C.~Pareige}, \bibinfo{author}{C.~Massey},
  \bibinfo{author}{P.~Zhu}, \bibinfo{author}{J.~D. Poplawsky},
  \bibinfo{author}{J.~Henry}, \bibinfo{author}{S.~J. Zinkle},
\newblock \bibinfo{title}{Effect of heavy ion irradiation dose rate and
  temperature on {$\alpha{'}$} precipitation in high purity {{Fe-18}}\%{{Cr}}
  alloy},
\newblock \bibinfo{journal}{Acta Materialia} \bibinfo{volume}{231}
  (\bibinfo{year}{2022}) \bibinfo{pages}{117888}.
  \DOIprefix\doi{10.1016/j.actamat.2022.117888}.
\bibitem[{Haley et~al.(2017)Haley, Briggs, Edmondson, Sridharan, Roberts,
  {Lozano-Perez}, and Field}]{haleyDislocationLoopEvolution2017}
\bibinfo{author}{J.~C. Haley}, \bibinfo{author}{S.~A. Briggs},
  \bibinfo{author}{P.~D. Edmondson}, \bibinfo{author}{K.~Sridharan},
  \bibinfo{author}{S.~G. Roberts}, \bibinfo{author}{S.~{Lozano-Perez}},
  \bibinfo{author}{K.~G. Field},
\newblock \bibinfo{title}{Dislocation loop evolution during in-situ ion
  irradiation of model {{FeCrAl}} alloys},
\newblock \bibinfo{journal}{Acta Materialia} \bibinfo{volume}{136}
  (\bibinfo{year}{2017}) \bibinfo{pages}{390--401}.
  \DOIprefix\doi{10.1016/j.actamat.2017.07.011}.
\bibitem[{Zhang et~al.(2021)Zhang, Xiao, and
  Bai}]{zhangEffectCrConcentration2021}
\bibinfo{author}{Y.~Zhang}, \bibinfo{author}{Z.~Xiao}, \bibinfo{author}{X.-M.
  Bai},
\newblock \bibinfo{title}{Effect of {{Cr Concentration}} on
  {$\frac{1}{2}<$}111{$>$} to {$<$}100{$>$} {{Dislocation Loop Transformation}}
  in {{Fe-Cr}} alloys},
\newblock \bibinfo{journal}{Journal of Nuclear Materials} \bibinfo{volume}{543}
  (\bibinfo{year}{2021}) \bibinfo{pages}{152592}.
  \DOIprefix\doi{10.1016/j.jnucmat.2020.152592}.
\bibitem[{Cottrell et~al.(2004)Cottrell, Dudarev, and
  Forrest}]{cottrellImmobilizationInterstitialLoops2004}
\bibinfo{author}{G.~A. Cottrell}, \bibinfo{author}{S.~L. Dudarev},
  \bibinfo{author}{R.~A. Forrest},
\newblock \bibinfo{title}{Immobilization of interstitial loops by
  substitutional alloy and transmutation atoms in irradiated metals},
\newblock \bibinfo{journal}{Journal of Nuclear Materials} \bibinfo{volume}{325}
  (\bibinfo{year}{2004}) \bibinfo{pages}{195--201}.
  \DOIprefix\doi{10.1016/j.jnucmat.2003.12.001}.
\bibitem[{Ukai et~al.(2021)Ukai, Yano, Inoue, and
  Sowa}]{ukaiSolidsolutionStrengtheningCr2021}
\bibinfo{author}{S.~Ukai}, \bibinfo{author}{Y.~Yano},
  \bibinfo{author}{T.~Inoue}, \bibinfo{author}{T.~Sowa},
\newblock \bibinfo{title}{Solid-solution strengthening by {{Al}} and {{Cr}} in
  {{FeCrAl}} oxide-dispersion-strengthened alloys},
\newblock \bibinfo{journal}{Materials Science and Engineering: A}
  \bibinfo{volume}{812} (\bibinfo{year}{2021}) \bibinfo{pages}{141076}.
  \DOIprefix\doi{10.1016/j.msea.2021.141076}.

\end{thebibliography}
  \biboptions{sort&compress}

%
%
%
\end{document}